\documentclass[floats,floatfix,amssymb,prd,twocolumn,nofootinbib]{revtex4-1}
\usepackage{amssymb,amsmath,verbatim,mathtools,needspace,enumitem,etoolbox,graphicx,physics,microtype}

\usepackage[dvipsnames, usenames]{xcolor}

\definecolor{linkcolor}{rgb}{0.0,0.3,0.5}
\definecolor{urlcolor}{rgb}{0.27,0.55,0.}
\definecolor{funcolor}{rgb}{0.65, 0.16, 0.16}
\usepackage[unicode, colorlinks=true, linkcolor=linkcolor, citecolor=linkcolor, filecolor=linkcolor,urlcolor=linkcolor, pdfusetitle]{hyperref}
\usepackage[all]{hypcap}

\usepackage[T1]{fontenc}
\usepackage[utf8]{inputenc}

\allowdisplaybreaks
\def\apj{\rm{ApJ}}
\def\apjl{\rm{ApJ}}

\def\aap{\rm{A\&A}}

\def\mnras{\rm{MNRAS}}
\def\prd{\rm{PRD}}
\def\prl{\rm{PRL}}
\def\prx{\rm{PRX}}

\def\cqg{\rm{CQG}}
\def\grg{\rm{GRG}}

\begin{document}

\title{Black-hole kicks from numerical-relativity surrogate models}

\author{Davide Gerosa}
\thanks{Einstein Fellow}
\email{dgerosa@caltech.edu}

\author{Fran\c{c}ois~H\'{e}bert}
\email{fhebert@caltech.edu}

\author{Leo C.~Stein}
\email{leostein@tapir.caltech.edu}

\affiliation{TAPIR 350-17, California Institute of Technology, 1200 E
  California Boulevard, Pasadena, CA 91125, USA}

\date{\today}

\begin{abstract}

Binary black holes radiate linear momentum in gravitational waves as
they merge. Recoils imparted to the black-hole remnant can reach
thousands of km/s, thus ejecting black holes from their host
galaxies. We exploit recent advances in gravitational waveform
modeling to quickly and reliably extract recoils imparted to generic,
precessing, black-hole binaries.
Our procedure uses a numerical-relativity surrogate model to obtain
the gravitational waveform given a set of binary parameters; then, from
this waveform we directly integrate the gravitational-wave linear
momentum flux.
This entirely bypasses the need for fitting formulas which are
typically used to model black-hole recoils in astrophysical
contexts. We provide a thorough exploration of the black-hole kick
phenomenology in the parameter space, summarizing and extending
previous numerical results on the topic.
Our extraction procedure is made publicly available as a module for
the Python programming language named \textsc{surrkick}. Kick
evaluations take $\sim 0.1$ s on a standard off-the-shelf machine,
thus making our code ideal to be ported to large-scale astrophysical
studies.%
\end{abstract}

\maketitle

\section{Introduction}

Gravitational waves (GWs) carry energy, linear momentum, and angular
momentum, and are therefore responsible for the final evolutionary
stages of compact binary systems. As energy and angular momentum are
dissipated away, the two objects inspiral and eventually merge. The
GW-driven orbital decay of two neutron stars was first observed
by pulsar timing, leading to a major confirmation of Einstein's theory
of general relativity~\cite{1982ApJ...253..908T}. The first landmark
detection of GWs was from a binary black hole (BH) which was brought
to merger by those same GWs that ultimately reached our
detectors~\cite{2016PhRvL.116f1102A}.

Similar to how the dissipation of energy and angular momentum causes the orbit
of a BH binary to shrink, the emission of linear momentum through GWs
causes the binary's center of mass to
recoil~\cite{1961RSPSA.265..109B, 1962PhRv..128.2471P}.
The key property to generate a GW recoil (or ``kick'') is
asymmetry. It is straightforward to show that symmetry prevents linear
momentum dissipation during the inspiral and merger of equal-mass,
nonspinning BHs. Conversely, a generic BH binary radiates GWs
anisotropically: linear momentum is preferentially emitted in some
direction, and the binary consequently recoils. BH kicks were first
studied using the post-Newtonian (PN) approximation
(e.g., Refs~\cite{1983MNRAS.203.1049F, 1995PhRvD..52..821K,
  2005ApJ...635..508B}) but their full astrophysical relevance was
only realized after numerical relativity (NR) simulations of BH
mergers became possible \cite{2005PhRvL..95l1101P,2006PhRvL..96k1101C,2006PhRvL..96k1102B}.  Most of the linear
momentum is emitted during the last few orbits and merger, which
corresponds to the highly dynamical, fully nonlinear regime that can
only be captured with NR simulations.

In particular, simulations showed that BHs formed following a merger
may be imparted recoil velocities of up to $5000$
km/s~\cite{2007PhRvL..98w1102C, 2007PhRvL..98w1101G,
  2007PhRvD..76f1502T, 2011PhRvL.107w1102L}.  The striking
astrophysical consequences of these findings were quickly realized
(e.g., Refs.~\cite{2007MNRAS.382L...6S, 2007ApJ...662L..63S,
  2008ApJ...678..780G, 2008ApJ...682..758S, 2008ApJ...686..829H,
  2008MNRAS.390.1311B}): BH recoils might exceed the escape speed of
even the most massive galaxies in the
Universe~\cite{2004ApJ...607L...9M, 2015MNRAS.446...38G}, thus making
galactic ejections a possible outcome of binary
mergers~\cite{1989ComAp..14..165R}.  Recoiling BHs might give rise to
a variety of electromagnetic signatures~\cite{2012AdAst2012E..14K}
---notably a kinematical offset of a set of broad emission lines---
which led to the identifications of a few observational
candidates~\cite{2008ApJ...678L..81K, 2012ApJ...752...49C,
  2014MNRAS.445.1558D, 2014MNRAS.445..515K, 2017A&A...600A..57C,
  2017ApJ...840...71K, 2017ApJ...851L..15K} (see
also Refs.~\cite{2014ApJ...795..146L, 2016MNRAS.455..484R,
  2016MNRAS.456..961B} for detection strategies). As the system
recoils, a Doppler shift of the emitted GWs can provide a possible
direct observational signature of BH kicks within the reach of future
space- and ground-based GW observatories~\cite{2016PhRvL.117a1101G}.

Since NR simulations are far too expensive to be performed in
astrophysical population studies, BH kicks have mostly been modeled
using fitting formulas based on PN theory and
calibrated to NR simulations (e.g., Refs.~\cite{2007ApJ...659L...5C,
  2007PhRvL..98i1101G, 2008PhRvD..77d4028L, 2008PhRvD..77d4028L,
  2012PhRvD..85h4015L, 2013PhRvD..87h4027L}).  These ``black box''
expressions return the final kick of the BH remnant given the
intrinsic parameters (mass ratio and spins) of the merging binary at
some initial separation. Another so far unexplored possibility to
model BH kicks is to compute the flux of linear momentum in GWs using
a waveform approximant that can be quickly evaluated in parameter
space. Linear momentum dissipation, however, is encoded in both
differences between the dominant $l=2, m=\pm 2$ modes and higher
harmonics ($l>2$)~\cite{2008PhRvD..77l4047B}. This approach,
therefore, requires an inspiral-merger-ringdown approximant able to
model both higher harmonics (crucial to linear momentum flux) and
misaligned spins (which are known to generate the largest kicks).

In this paper we present the first attempt in this direction using the
recent NR surrogate model by Blackman \emph{et
  al.}~\cite{2017PhRvD..96b4058B} --- the first waveform approximant
able to model generic precessing systems with higher harmonics. In
contrast with the available fitting formulas, our procedure provides
not only the final kick speed $v_k$, but also the entire velocity
accumulation profile $\mathbf{v}(t)$.  We present a thorough
exploration of BH recoils for generic systems, which summarizes and
extends various previous findings in a coherent fashion.
Our numerical code, \textsc{surrkick},
is publicly available and allows for reliable computation of the
radiated quantities (energy, linear momentum, and angular momentum) at
a
moderate
computational cost. Our implementation is therefore ideal to be ported
to larger-scale astrophysical codes which require fast estimates of BH
kicks, such as galaxy merger-tree simulations, populations synthesis
studies, and GW event-rate predictions.

This paper is organized as follows. Section \ref{methods} introduces the
main tools of our analysis. Section \ref{results}  presents results and
comparisons with other methods. Section \ref{accuracy} explores the
numerical accuracy of our procedure. Section \ref{code} briefly describes the implementation and usage of our public code. Section \ref{conclusions}
draws conclusions and future prospects. Unless otherwise stated, we
use relativists' units $c=G=1$.

\section{Methods}
\label{methods}

\subsection{Numerical-relativity surrogate models}
\label{surrogatemodels}

Surrogate models interpolate a set of precomputed  GW signals and make
use of advanced decomposition and interpolation schemes to quickly
produce waveforms for any desired point in parameter space.  Surrogate
models are typically optimized to accurately reproduce the complex
gravitational-wave strain,
here expanded in terms of spin-weighted spherical
harmonics~\cite{1980RvMP...52..299T}
\begin{align}
h(t,\theta,\phi,\boldsymbol{\lambda}) &= h_+(t,\theta,\phi,\boldsymbol{\lambda}) - i h_\times(t,\theta,\phi,\boldsymbol{\lambda})
\notag \\
&=\sum_{l=2}^\infty \sum_{m=-l}^{+l} h^{lm}(t,\boldsymbol\lambda) \; _{-2}Y_{lm}(\theta,\phi)\,,
\label{hmodes}
\end{align}
where $t$ denotes time, $\theta$ and $\phi$ describe the GW
propagation direction, and the symbol $\boldsymbol\lambda$ encodes all
the binary's intrinsic parameters. For quasicircular BH binaries,
these are  the mass ratio $q$ and spin vectors
$\boldsymbol{\chi_1},\boldsymbol{\chi_2}$ (the total mass $M$ is a
free scale).

Surrogate models have been presented for both
effective-one-body~\cite{2014PhRvX...4c1006F, 2014CQGra..31s5010P,
  2016PhRvD..93f4041P} and NR waveforms~\cite{2017PhRvD..95j4023B,
  2017PhRvD..95j4023B, 2017PhRvD..96b4058B}.
In this paper we use the NR waveform surrogate model
\mbox{NRSur7dq2}~\cite{2017PhRvD..96b4058B} to generate our
waveforms. \mbox{NRSur7dq2} is the very first model able to cover the
seven-dimensional parameter space describing generic precessing
systems. \mbox{NRSur7dq2} is trained on 886 NR waveforms generated
with the Spectral Einstein Code (SpEC)~\cite{2000PhRvD..62h4032K}
and interpolated using the technique put forward
in Ref.~\cite{2014PhRvX...4c1006F}. It provides modes $h^{lm}$ up to
$l\leq4$ for binaries with mass ratios $q=m_2/m_1 \in[0.5,1]$ and
dimensionless spin magnitudes $\chi_1,\chi_2\in[0, 0.8]$; updates to
extend its validity range are under active development. The model has
been shown to be extremely accurate at reproducing the
gravitational-wave strain $h$: it outperforms all other available
waveform approximants by several orders of magnitude, reaching a level
of accuracy comparable to the NR simulations used in the training
process~\cite{2017PhRvD..96b4058B}.

Waveforms generated with \mbox{NRSur7dq2} span the %
time range $-4500M \leq t \leq 100M$, where $t=0$ is defined as the
time that maximizes the total waveform amplitude
$\mathcal{A}^2(t) = \sum_{l,m} | h^{lm}(t) |^2$.  The initial time
$t=-4500M$ corresponds to about $20$ orbits before merger and the
final value $t=100M$ allows for a full dissipation of the signal.
Values of $h^{lm}$ are computed at carefully selected time
nodes~\cite{2017PhRvD..96b4058B} and later interpolated
in time
using standard cubic univariate B-splines.  More specifically,
\mbox{NRSur7dq2} provides the distance-independent dimensionless
strain, extrapolated to $\mathcal{I}^+$, i.e.~$\lim_{r\to \infty} r h
/ M$ where $r$ is the distance from the binary's center of mass and
$M$ is the total mass of the binary at the beginning of the
evolution.
\mbox{NRSur7dq2} allows for the spin directions to be specified at a
reference time $-4500M\leq t_{\rm ref}\leq -100M$, in a  frame defined
such that the more (less) massive BH sits on the positive (negative)
x-axis and the  Newtonian orbital angular momentum $\mathbf{L}$ lies
along the z-axis. Unless otherwise stated, we use $t_{\rm ref}=-100M$.

\subsection{Radiated energy and momenta}
\label{radiatedexpressions}
Multipolar expansions for the radiated energy, linear momentum and
angular momentum have been worked out in detail
in~Ref.~\cite{2008GReGr..40.1705R} (derived
from Refs.~\cite{1980RvMP...52..299T, 2007PhRvD..76d1502L}).
We report their expressions here for completeness.\footnote{%
  The author of Ref.~\cite{1980RvMP...52..299T} presented his formulas in
  specially chosen coordinate systems.  A more rigorous mathematical
  framework for these calculations is to go to $\mathcal{I}^{+}$ and
  present the news tensor, Bondi mass aspect, and other Bondi charges
  (e.g.~Ref.~\cite{2017PhRvD..95d4002F}). The authors of Ref.~\cite{2008GReGr..40.1705R}
  used the convention $\Im(a+ib)=ib$, while here we use $\Im(a+ib)=b$.}
Whenever terms with $l<2$ or $|m|>l$ are present in the following
summations, their coefficients are intended to be zero. In practice,
one is also limited to $l\leq l_{\rm max}$ (where, e.g., $l_{\rm max}=4$
for \mbox{NRSur7dq2} waveforms and $l_{\rm max}=8$ for SpEC
waveforms). %

The energy flux emitted in GWs is provided in terms of the first time
derivative of the complex strain $\dot h$ and reads:
\begin{align}
\frac{dE}{dt} = \lim_{r \rightarrow \infty} \frac{r^2}{16\,\pi}
\sum_{l,m} \,\left| \dot h^{l,m} \right|^2 \; .
\label{energyflux}
\end{align}
When integrating to obtain $E(t)$  we set the integration constant
$E_0$ to account for the binding energy dissipated in GWs at times
$t<-4500M$, before the start of our waveforms, thus enforcing
$\lim_{t\to-\infty} E(t)=0$. A straightforward Newtonian calculation yields~\cite{1964PhRv..136.1224P}
\begin{equation}
\frac{E_0}{M}= \left( \frac{5}{1024} \frac{q^3}{(1+q)^6} \dot E_0\right)^{1/5},
\end{equation}
where $\dot E_0$ is estimated from Eq.~(\ref{energyflux}) by averaging
over the first $100M$ in time.  We have verified that corrections up
to 2PN (including spin effects~\cite{Arun:2008kb}) have a negligible
impact on $E_0$.  One can then define the time-dependent (Bondi) mass
of the binary,
\begin{equation}
M(t) = M - E(t) + E_0
\,,
\label{Moft}
\end{equation}
such that $M(t)$ at the beginning of our waveforms is equal to
$M$. The mass of the post-merger BH in units of the total mass of the
binary at early times is
\begin{equation}
\frac{ \displaystyle\lim_{t\to+\infty} M(t)}{{\displaystyle \lim_{t\to-\infty} M(t)}} =  1- \frac{\displaystyle \lim_{t\to+\infty} E(t)}{M+E_0}.
\label{masslimits}
\end{equation}

The emitted linear momentum is also fully specified by  $\dot h$ and
crucially includes mixing between modes with different $l$ and $m$:
\begin{align}
\frac{d P_x}{dt} = &\lim_{r \to \infty} \frac{r^2}{8\, \pi} \Re \Bigg[ \sum_{l,m} \,
\dot h^{l,m}  \Big( a_{l,m}\, \dot{\bar{h}}^{l,m+1}
 \notag\\
&+ b_{l,-m} \,\dot{\bar{h}}^{l-1,m+1}  -  b_{l+1,m+1}\, \dot{\bar{h}}^{l+1,m+1} \Big)\Bigg] \; ,
\label{eq:dt_px} \\
\frac{d P_y}{dt} = &\lim_{r \to \infty} \frac{r^2}{8\, \pi}\Im \Bigg[  \sum_{l,m}\, \dot h^{l,m}  \Big( a_{l,m}\, \dot{\bar{h}}^{l,m+1}
 \notag\\
&+ b_{l,-m} \,\dot{\bar{h}}^{l-1,m+1}  -  b_{l+1,m+1}\, \dot{\bar{h}}^{l+1,m+1} \Big)\Bigg] \; ,
\label{eq:dt_py} \\
\frac{d P_z}{dt} = &\lim_{r \to \infty} \frac{r^2}{16 \pi} \sum_{l,m}\,
\dot{{h}}^{l,m}   \Big( c_{l,m}\, \dot{\bar{h}}^{l,m}
\notag\\
&+ d_{l,m}\, \dot{\bar{h}}^{l-1,m} +  d_{l+1,m}\, \dot{\bar{h}}^{l+1,m} \Big) \; ,
\label{eq:dt_pz}
\end{align}
where the upper bar denotes complex conjugation and
\begin{eqnarray}
a_{l,m} &=& \frac{\sqrt{(l-m)\,(l+m+1)}}{l\,(l+1)} \; , \\
b_{l,m} &=& \frac{1}{2\,l}\, \sqrt{\frac{(l-2)\,(l+2)\,(l+m)\,(l+m-1)}
{(2l-1)(2l+1)}} \; , \\
c_{l,m} &=& \frac{2\,m}{l\,(l+1)} \; , \\
d_{l,m} &=& \frac{1}{l}\, \sqrt{\frac{(l-2)\,(l+2)\,(l-m)\,(l+m)}
{(2l-1)(2l+1)}} \; .
\end{eqnarray}
The integration constant for the $d\mathbf{P}/dt$ integration is
chosen so that the average of $\mathbf{P}$ over the first $1000M$ in
time, where linear momentum emission is expected to be negligible, is
zero.
By conservation of linear momentum, the time profile of the kick
imparted to the system is\footnote{%
  Relativistic corrections are irrelevant here. The largest BH kicks
  are $v_k/c\sim 10^{-2}$, corresponding to Lorentz factors
  $\gamma-1 \sim 10^{-4}$.}
\begin{equation}
\mathbf{v}(t) = - \frac{{P_x}(t) \mathbf{\hat x} + {P_y}(t) \mathbf{\hat y} + {P_z}(t) \mathbf{\hat z}}{M(t)}\,,
\label{voftprofile}
\end{equation}
and the final velocity of the post-merger remnant BH is
\begin{align}
\mathbf{v_k} = \lim_{t\to \infty} \mathbf{v}(t)\,.
\label{vkicklimit}
\end{align}
One can further integrate $\mathbf{v}(t)$ in time to obtain the
trajectory $\mathbf{x}(t)= \int \mathbf{v}(t) dt$. Although the binary
trajectory is a coordinate-dependent notion, the time integral of the
linear momentum dissipated in GWs can be interpreted as the motion of
the spacetime's center of mass seen by an observer at
$\mathcal{I}^+$~\cite{2017PhRvD..95d4002F}.

The angular momentum carried by GWs involves both $h$ and $\dot h$:
\begin{align}
\frac{d J_x}{dt} =  &\lim_{r\rightarrow\infty} \frac{r^2}{32 \pi} \:
\Im \Bigg[ \sum_{l,m} \,h^{l,m}
\Big( f_{l,m}\, \dot{\bar{h}}^{l,m+1}
\notag \\&+ f_{l,-m}\, \dot{\bar{h}}^{l,m-1} \Big) \Bigg]\;  ,
\label{eq:dt_jx} \\
\frac{d J_y}{dt} = &- \lim_{r\rightarrow\infty} \frac{r^2}{32 \pi} \:
\Re \Bigg[ \sum_{l,m} \, h^{l,m}  \Big( f_{l,m}\, \dot{\bar{h}}^{l,m+1}
\notag \\ &- f_{l,-m}\, \dot{\bar{h}}^{l,m-1} \Big) \Bigg]\; ,
\label{eq:dt_jy} \\
\frac{d J_z}{dt} =  &\lim_{r\rightarrow\infty} \frac{r^2}{16 \pi} \:
\Im \Bigg[ \sum_{l,m} \,m\, h^{l,m}
 \,\dot{\bar{h}}^{l,m} \Bigg] \; ,
\label{eq:dt_jz}
\end{align}
where
\begin{eqnarray}
f_{l,m} = \sqrt{l(l+1) - m(m+1)} \; .
\label{flm}
\end{eqnarray}
When integrating $d\mathbf{J}/dt$, we do not adjust the integration
constant to account for the angular momentum radiated before the
beginning of our waveforms.  Contrary to the binding energy, the
Newtonian angular momentum of a binary system diverges as separation
grows ($J\propto \sqrt{r}$).

We perform all differentiations and integrations required to extract
these radiated quantities analytically on the spline interpolants
provided by \mbox{NRSur7dq2}, over the range $-4500M\leq t \leq
100M$.
The $t\to\infty$ limits [e.g.~Eqs.~(\ref{masslimits}) and
(\ref{vkicklimit})] are approximated with values at $t=100M$.

\section{Results}
\label{results}

\subsection{Anatomy of the kick}
\label{anatomy}

\begin{figure}[t]
\includegraphics[width=\columnwidth]{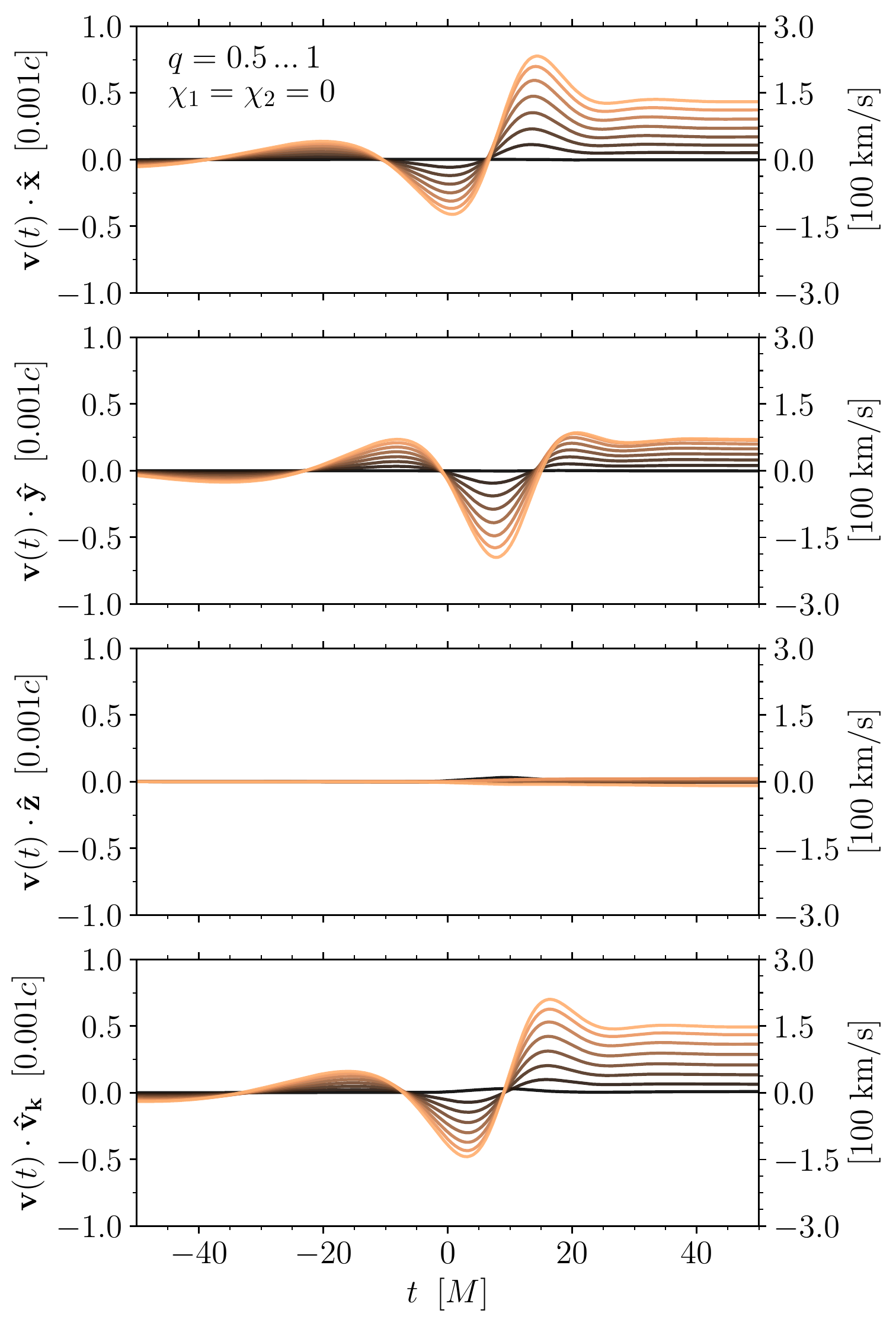}
\caption{Kick profile $\mathbf{v}(t)$ projected along
  $\mathbf{\hat x}$, $\mathbf{\hat y}$, $\mathbf{\hat z}$ and the
  direction of the final kick $\mathbf{\hat v_k}$ for a series of
  non-spinning BH binaries with mass ratio ranging from $q=0.5$ (light
  orange) to $q=1$ (black). The binary's center of mass oscillates in
  the orbital plane during the inspiral; the final recoil is imparted
  with a sudden acceleration at $t\sim 10M$ after the peak-amplitude
  time.}
\label{nospinprofiles}
\end{figure}

\begin{figure*}[t]
\includegraphics[page=1,clip,trim=1.cm 0cm 0cm 0cm, width=0.325\textwidth]{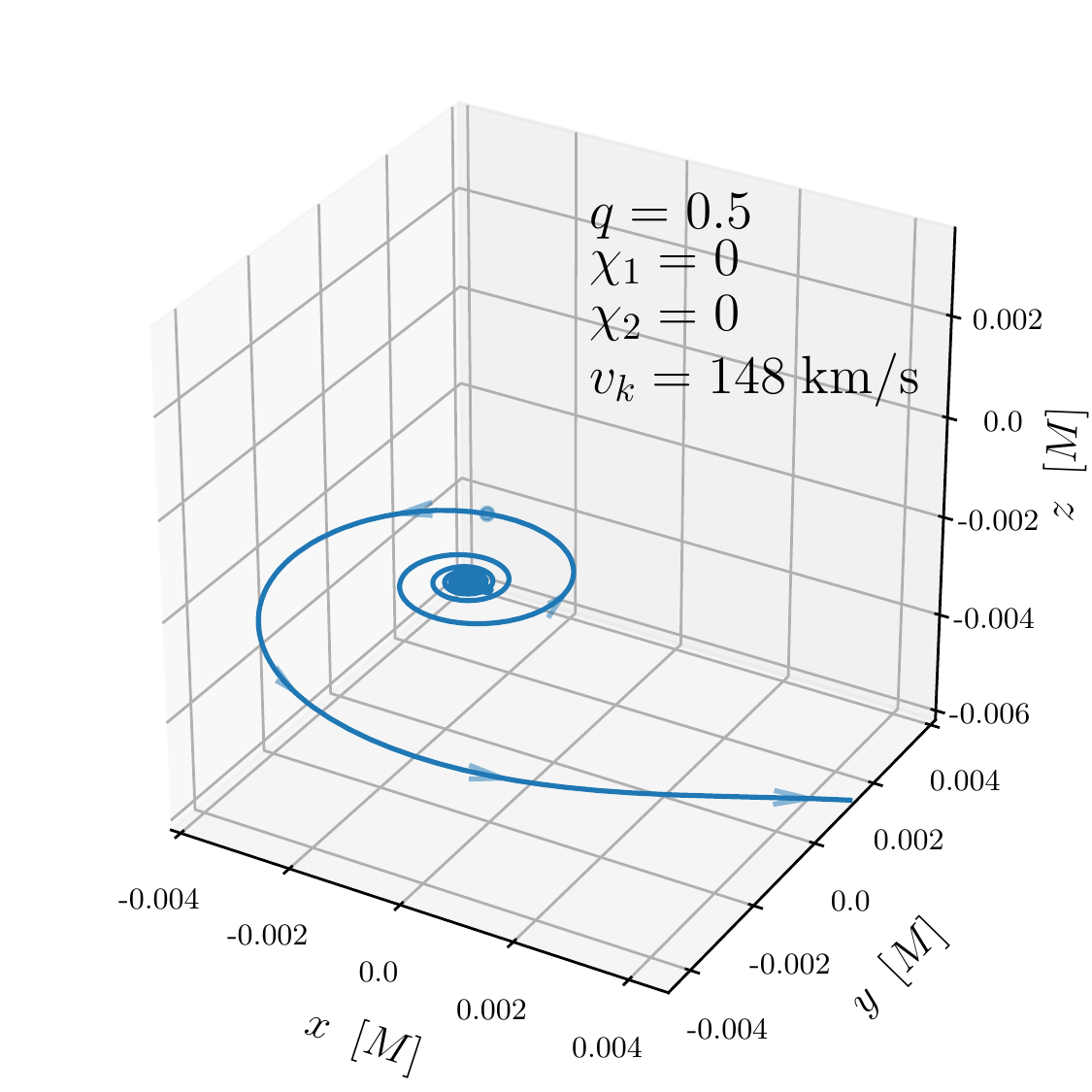}
\includegraphics[page=2,clip,trim=1.cm 0cm 0cm 0cm, width=0.325\textwidth]{centerofmass}
\includegraphics[page=3,clip,trim=1.cm 0cm 0cm 0cm, width=0.325\textwidth]{centerofmass}
\caption{Center-of-mass trajectory $\mathbf{x}(t)=\int\mathbf{v}(t)
  dt$ for three binary configurations as described in the legends. The
  circle markers on each curve correspond to $t=0$. The left panel
  shows a recoil due to mass asymmetry only: the center of mass
  oscillates in the orbital plane during the inspiral and is finally
  pushed after merger. The middle panel shows a complicated interplay
  of mass and spin asymmetry, with the initial oscillations being
  greatly distorted at merger by the superkick effect. Finally, the
  right panel shows the simpler trajectory of a binary receiving a
  very large kick of $\sim 3000$ km/s. An animated version of this
  figure is available at
  \href{https://davidegerosa.com/surrkick}{davidegerosa.com/surrkick}.}
\label{centerofmass}
\end{figure*}

Nonspinning BH binaries do not receive any recoil for both $q=1$
(because of symmetry) and $q=0$ (which corresponds to the
test-particle limit).  Recoils are present in between these two
limits.  Figure~\ref{nospinprofiles} shows the kick profile
$\mathbf{v}(t)$
for a series of BH mergers with $q=0.5,\dots, 1$. Axisymmetry prevents
linear momentum dissipation along the direction of the orbital angular
momentum, i.e.~$\mathbf{v}(t)\cdot \mathbf{\hat z}= 0$ (within
numerical errors; see Sec.~\ref{exploiting}). The binary's center of
mass oscillates in the orbital plane x-y during the inspiral, until the
merger halts these oscillations and imparts the final recoil.  The
kick velocity grows as $q$ decreases, reaching %
$v_k\simeq 148$ km/s for $q=0.5$. The largest kick achievable for a
nonspinning system is $v_k\simeq 175$ km/s and corresponds to
$q\sim 0.36$~\cite{2007PhRvL..98i1101G}, which is outside the
parameter space currently covered by \mbox{NRSur7dq2}. The trajectory
of the spacetime's center of mass $\mathbf{x}(t)$ for $q=0.5$ and
$\chi_1=\chi_2=0$ is shown in the left panel of
Fig.~\ref{centerofmass}. One last oscillation occurs after merger, and
is responsible for most of the kick.
This effect is also visible in Fig.~\ref{nospinprofiles}, where we see
the system typically accelerates at $t\sim 10M$ after merger, with the
final burst of linear momentum radiation lasting only for a few $M$ in
time. Interestingly, the projection of the recoil profile along the
final kick direction $\mathbf{v}(t)\cdot \mathbf{\hat v_k}$ is not
monotonic after merger: the binary suddenly decelerates at about
$t\sim 15M$, after which the imparted velocity settles down to the
asymptotic value $v_k$. This effect has been dubbed
\emph{antikick}~\cite{2010PhRvL.104v1101R}, and turns out to be a
rather generic feature of BH mergers (cf.~Sec.~\ref{statistics}
below).

\begin{figure}[t]
\includegraphics[width=\columnwidth]{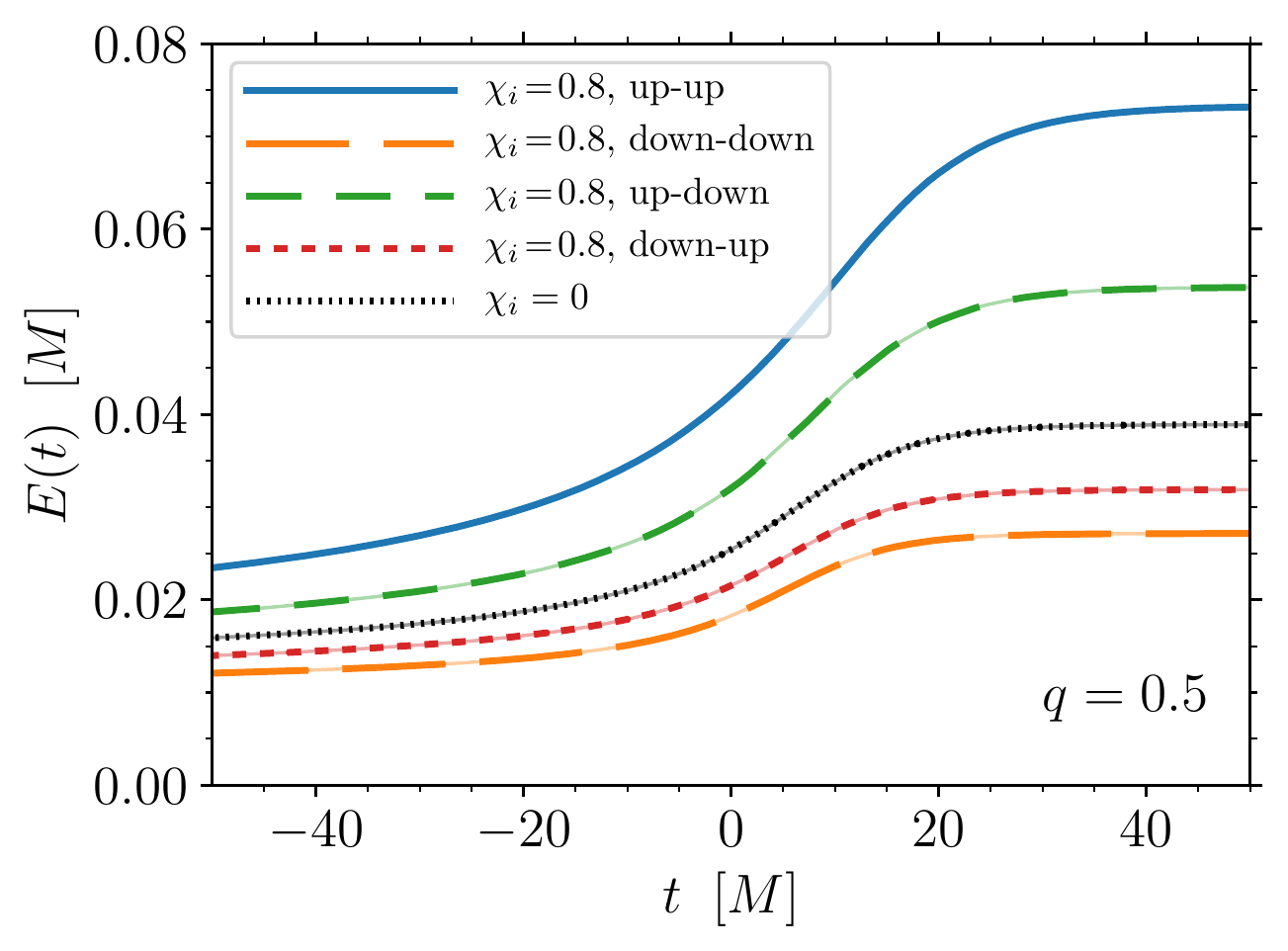}
\caption{Radiated energy ${E}(t)$ for binaries with mass ratio $q=0.5$
  and spins of magnitude $\chi_1=\chi_2=0.8$ (anti)aligned to the
  orbital angular momentum. Four configurations are shown ---up-up,
  down-down, up-down, down-up--- where the term before (after) the
  hyphen refers to the spin of the heavier (lighter) BH being
  co-/counter-aligned with the binary's orbital angular momentum.  For
  comparison, we also show $E(t)$ for a non-spinning system with the
  same mass ratio. Because of the orbital hang-up effect, BH binaries
  with (anti-)aligned spins radiate more (less) energy compared to
  non-spinning systems with the same mass ratio.}
\label{hangupErad}
\end{figure}
\begin{figure*}
\centering
\includegraphics[page=1,width=0.47\textwidth]{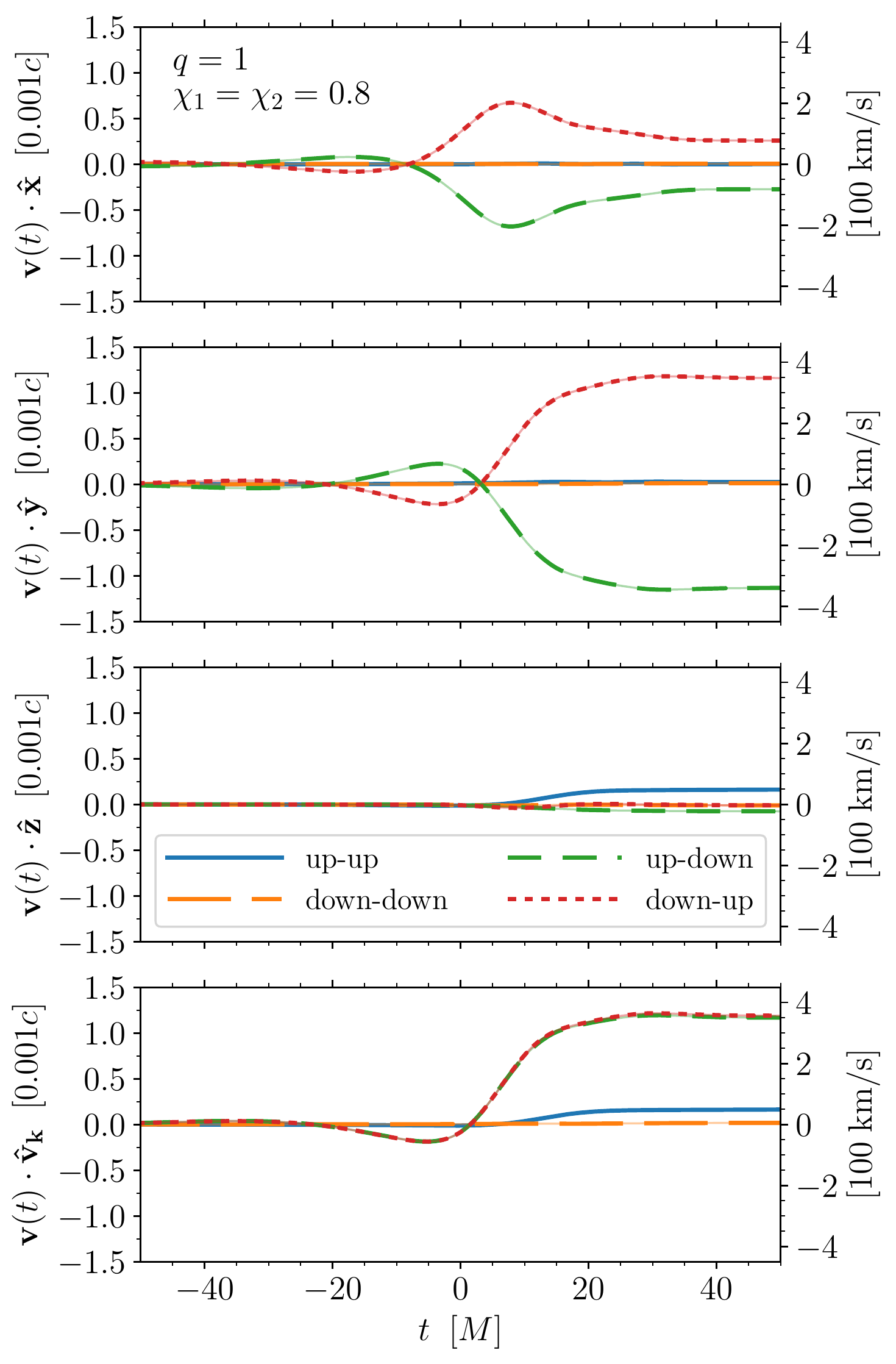}
\hspace{0.04\textwidth}
\includegraphics[page=2,width=0.47\textwidth]{spinaligned}
\caption{Kick profile $\mathbf{v}(t)$ projected along  $\mathbf{\hat
    x}$, $\mathbf{\hat y}$, $\mathbf{\hat z}$ and the direction of the
  final kick $\mathbf{\hat v_k}$ for binaries with mass ratio $q=1$
  (left) and $q=0.5$ (right), and spins of magnitude
  $\chi_1=\chi_2=0.8$ (anti)aligned to the orbital angular
  momentum. Four configurations are shown: up-up, down-down, up-down,
  down-up, where the term before (after) the hyphen refers to the spin
  of the heavier (lighter) BH being co-/counter-aligned with the
  binary's orbital angular momentum.  Kicks from non-precessing
  systems lie in the binary's orbital plane, with the spin kicks being
  more pronounced for the up-down and down-up configurations in
  accordance with PN predictions.}
\label{spinaligned}
\end{figure*}

BH spins introduce additional sources of linear momentum dissipation.
The impact of aligned spins on the radiated energy and linear momentum
profile is illustrated in Figs.~\ref{hangupErad} and \ref{spinaligned},
respectively.  In particular, we study BH binaries with spin magnitude
$\chi_1=\chi_2=0.8$ and four different spin orientations:
$\boldsymbol{\hat \chi_1}\cdot \mathbf{\hat z}=\boldsymbol{\hat \chi_2}\cdot \mathbf{\hat z}=1$ (up-up),
$\boldsymbol{\hat \chi_1}\cdot \mathbf{\hat z}=\boldsymbol{\hat \chi_2}\cdot \mathbf{\hat z}=-1$ (down-down),
$\boldsymbol{\hat \chi_1}\cdot \mathbf{\hat z}=-\boldsymbol{\hat \chi_2}\cdot \mathbf{\hat z}=1$ (up-down),
$\boldsymbol{\hat \chi_1}\cdot \mathbf{\hat z}=-\boldsymbol{\hat \chi_2}\cdot \mathbf{\hat z}=-1$ (down-up), where $\mathbf{\hat z} =\mathbf{\hat L}$ at $t_{\rm ref}=-100M$. Although the up-down configuration is generically unstable to spin precession~\cite{2015PhRvL.115n1102G}, the instability develops on longer timescales and can therefore be neglected in this context.  The orbital hang-up effect~\cite{2001PhRvD..64l4013D, 2006PhRvD..74d1501C, 2015CQGra..32j5009S} causes binaries with spins co- (counter-) aligned with the binary's angular momentum to merge later (sooner) compared to non-spinning systems with the same mass ratio. Consequently, the energy emitted in GWs  increases (decreases) if the total spin $\mathbf{S} = m_1^2 \boldsymbol{\chi_1} + m_2^2\boldsymbol{\chi_2}$ is (\mbox{anti-})aligned with $\mathbf{L}$ (c.f.~Fig.~\ref{hangupErad}).
For $q=1$ (Fig.~\ref{spinaligned}, left panel), moderately large
recoils of $v_k\sim 350$ km/s are achieved for the up-down and down-up
configurations, in agreement with the PN predictions $v_k\propto
|\boldsymbol{\hat \chi_1}\cdot \mathbf{\hat L}-\boldsymbol{\hat
  \chi_2}\cdot \mathbf{\hat L}|$~\cite{1995PhRvD..52..821K}
(see~\cite{2007ApJ...668.1140B,2008PhRvD..77d4028L} for numerical
explorations).
The recoil is mostly imparted in the orbital plane, but its magnitude
is somewhat smaller than the mass-asymmetry case explored above and
reduces to a single burst of linear momentum emitted at $t\sim10M$,
preceded by a
smaller one in the opposite direction at $t\sim-5M$.
The $q=1$ up-up configuration presents some linear momentum emitted
perpendicular to the orbital plane, resulting in $v_k \sim 50$
km/s. This is the inherent error scale in our model,
as symmetry implies $v_k=0$ for both the up-up and down-down
configuration at $q=1$~\cite{2008PhRvL.100o1101B,2008PhRvD..78b4017B},
see Sec.~\ref{exploiting}.  For binaries with unequal masses and
aligned spins (Fig.~\ref{spinaligned}, right panel), both the orbital
hang-up and the mass asymmetry effect are present: the binary's center
of mass first oscillates in the orbital plane (%
because
 $q\neq 1$) and then receive a further push at $t\sim10M$ (%
because
$\boldsymbol{\chi_i} \cdot \mathbf{\hat z}\neq 0$).

\begin{figure*}
\includegraphics[page=1,width=0.47\textwidth]{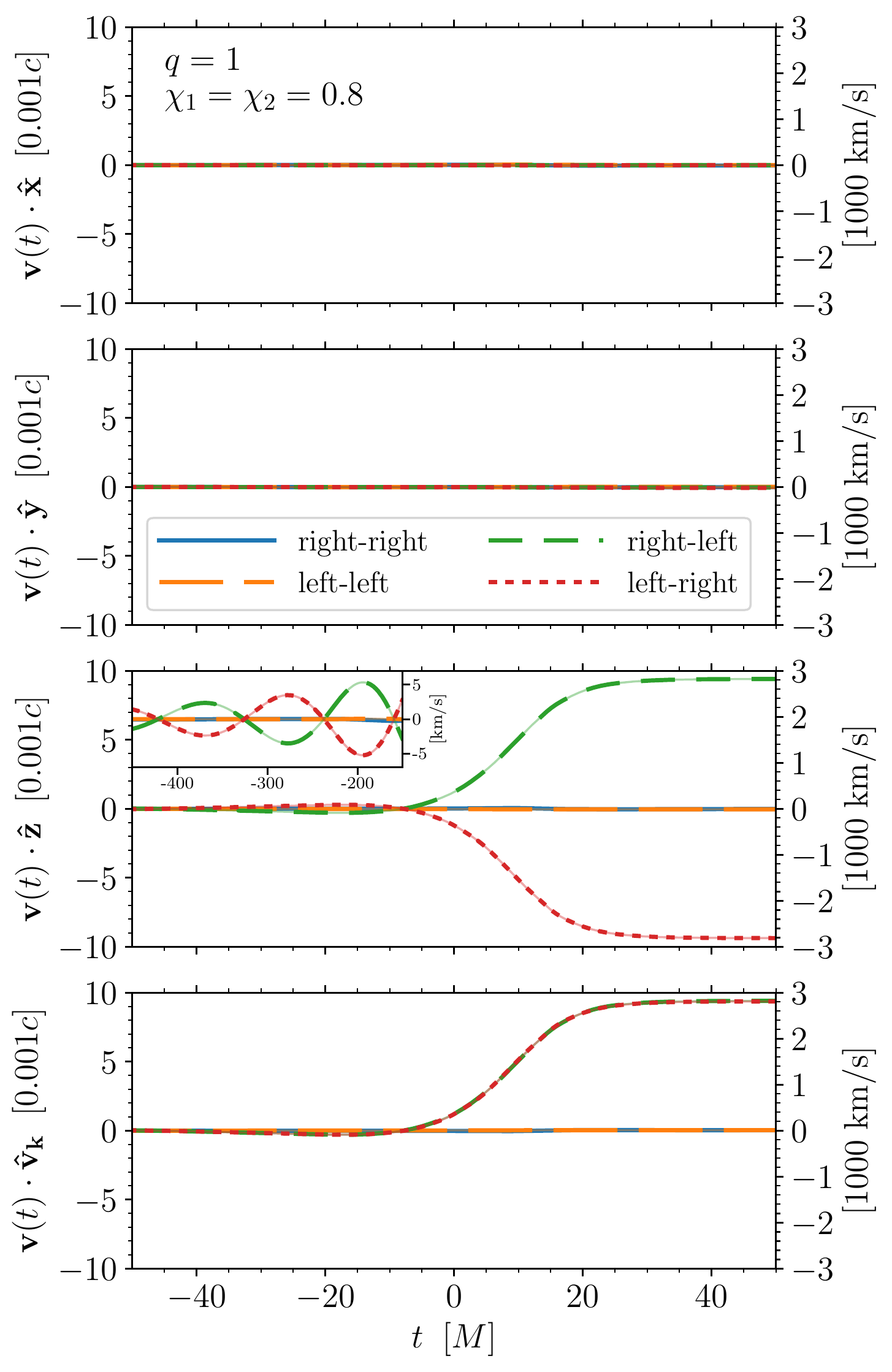}
\hspace{0.04\textwidth}
\includegraphics[page=2,width=0.47\textwidth]{leftright}
\caption{Kick profile $\mathbf{v}(t)$ projected along $\mathbf{\hat
    x}$, $\mathbf{\hat y}$, $\mathbf{\hat z}$ and the direction of the
  final kick $\mathbf{\hat v_k}$ for binaries  $q=1$ (left) and
  $q=0.5$ (right), and spins of magnitude $\chi_1=\chi_2=0.8$ lying
  into the orbital plane.  Four configurations are shown: right-right,
  left-left, right-left, left-right, where the term before (after) the
  hyphen refers to the spin of the heavier (lighter) BH being
  co-/counter-aligned with  initial separation vector $\mathbf{\hat
    x}$. The right-left and left-right  orientations correspond to the
  superkick configurations. Here we set $t_{\rm ref}=-125M$ to
  maximize kicks for the $q=1$ case (c.f.~Fig.~\ref{alphaseries}).}
\label{leftright}
\end{figure*}

\begin{figure*}
\includegraphics[width=\textwidth]{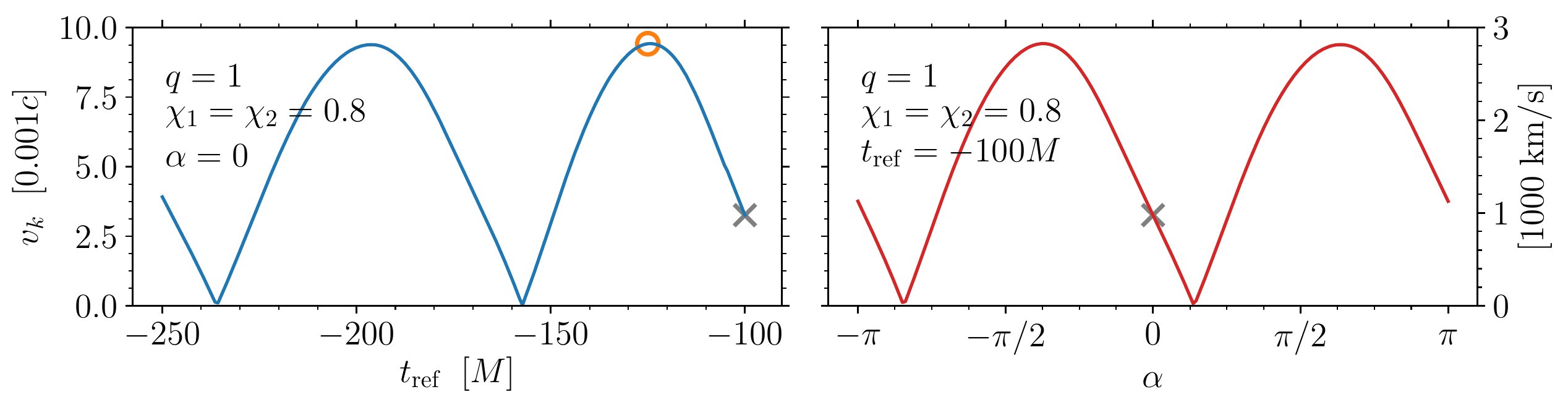}
\caption{Left panel: Recoil velocities for a series of right-left
  binaries with $q=1$ and $\chi_i=0.8$ initialized at various
  reference times $t_{\rm ref}$; the orange circle marks the reference
  time used in Fig.~\ref{leftright}. Right panel: Recoil velocities
  for BH binaries with $q=1$ and
  $\boldsymbol{\chi_1}=-\boldsymbol{\chi_2}=[0.8
  \cos\alpha,0.8\sin\alpha,0]$ (such that $\alpha=0$ corresponds to
  the right-left configuration) at $t_{\rm ref}=-100M$. The angle
  $\alpha$ corresponds to a rotation of both spins about the orbital
  angular momentum, and is degenerate with the reference time at which
  spins are specified. Gray crosses mark the same configuration in
  both panels.}
\label{alphaseries}
\end{figure*}

The largest kicks are achieved for BHs merging with misaligned
spins~\cite{2007PhRvL..98w1102C, 2007PhRvL..98w1101G,
  2007ApJ...659L...5C, 2007PhRvD..76f1502T, 2008PhRvD..77l4047B,
  2011PhRvL.107w1102L}.  Figure~\ref{leftright} shows kick profiles for
four binary configurations with spins $\chi_i=0.8$ lying in the
orbital plane:
$\boldsymbol{\hat \chi_1}\cdot \mathbf{\hat x}=\boldsymbol{\hat \chi_2}\cdot \mathbf{\hat x}=1$ (right-right),
$\boldsymbol{\hat \chi_1}\cdot \mathbf{\hat x}=\boldsymbol{\hat \chi_2}\cdot \mathbf{\hat x}=-1$ (left-left),
$\boldsymbol{\hat \chi_1}\cdot \mathbf{\hat x}=-\boldsymbol{\hat \chi_2}\cdot \mathbf{\hat x}=1$ (right-left),
$\boldsymbol{\hat \chi_1}\cdot \mathbf{\hat x}=-\boldsymbol{\hat \chi_2}\cdot \mathbf{\hat x}=-1$ (left-right),
where $\mathbf{\hat x}$ is defined as the axis connecting the lighter
to the heavier BH at $t_{\rm ref}$. For reasons clarified below, here
we take $t_{\rm ref} = -125 M$.  Kicks as large as $\sim 2820$ km/s
are achieved for the right-left and left-right configurations, which
correspond to the \emph{superkick} scenario discovered
in Refs.~\cite{2007PhRvL..98w1102C,2007PhRvL..98w1101G}. During the
inspiral, frame dragging from the two holes acts constructively and
pushes the binary's center of mass up and down along the direction of
the orbital angular momentum $\mathbf{\hat z}$. The final kick is
imparted as the BHs merge and the
last of these oscillations is abruptly interrupted. The phenomenology
is rather similar to the case of aligned spins studied above, although
with the key difference that in this case linear momentum is emitted
along the binary's orbital angular momentum, not orthogonal to it. It
is worth noting that binaries with these large kicks present a
remarkably simple accumulation profile: the acceleration
$d\mathbf{P}/dt$ is well described by a Gaussian centered at $t\sim
10M$ with width $\sigma\sim 5M$ (cf.~\cite{2008PhRvD..77l4047B} and
Sec.~\ref{statistics} below).
Conversely, frame dragging from the two BHs add destructively for the
right-right and left-left binaries.
This cancellation is perfect (within numerical errors,
cf.~Sec.~\ref{exploiting}) if the two spins have the same magnitude
$m_1^2 \chi_1 = m_2^2 \chi_2$ (Fig.~\ref{leftright}, left panel).
For $q=0.5$ and $\chi_i=0.8$ (Fig.~\ref{leftright}, right panel), the
dynamics is dominated by the largest spin and the four configurations
reach values between 650 and 1530 km/s.
Interestingly, smaller mass ratios excite a sizable kick along the
orbital plane of $\sim 300$ km/s, which exceed the recoil imparted to
nonspinning systems with the same $q$ of about  a factor $\sim 2$
(cf.~Fig.~\ref{nospinprofiles}).  The spacetime trajectory
$\int\mathbf{v}(t) dt$ for one such binary is illustrated in the
middle panel of Fig.~\ref{centerofmass}: the center of mass oscillates
at early time, undergoes a complicated motion right before merger,
after which the superkick effect becomes dominant. To the best of our
knowledge, this mass-spin asymmetry mixing in the kick profile has not
been reported elsewhere.

\begin{figure}
\includegraphics[width=0.95\columnwidth]{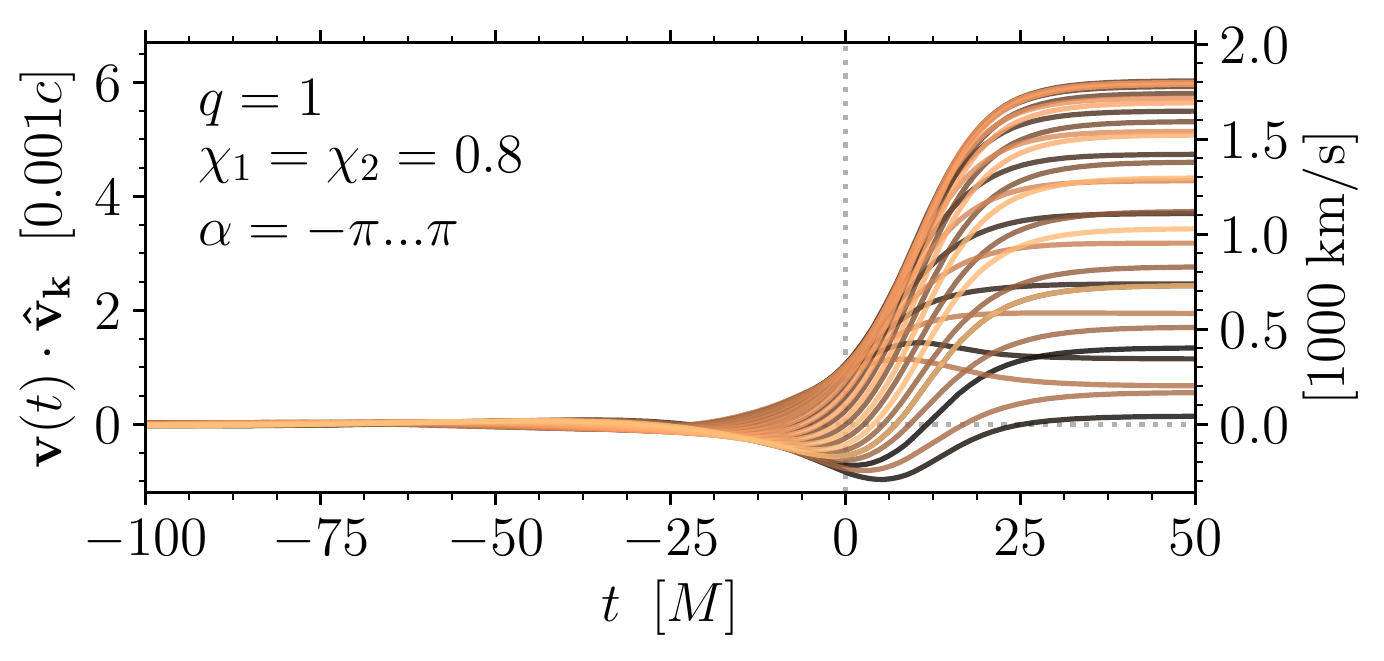}
\caption{Velocity accumulation profile $\mathbf{v}(t)$ projected along
  the direction of the final kick $\mathbf{\hat v_k}$ for binaries
  with $q=1$ and antiparallel spins of magnitude $\chi_1=\chi_2=0.8$
  lying in the orbital plane. The rotation angle $\alpha$ (defined as
  $\cos\alpha=\boldsymbol{\hat \chi_1}\cdot \mathbf{\hat x}=
  -\boldsymbol{\hat \chi_2}\cdot \mathbf{\hat x}$) controls the
  orbital phase at merger and thus sets the velocity of the center of
  mass  when the final kick is imparted. Curves are colored according
  to $\alpha$ as it spans from $-\pi$ (black) to $\pi$ (orange).}
\label{alphaprof}
\end{figure}

Superkick velocities critically depend on the orbital phase at merger,
as it controls the abrupt interruption of the oscillatory behavior
described above.
In the left panel of Fig.~\ref{alphaseries} we study a series of
right-left binaries ($q=1$, $\chi_1=\chi_2=0.8$, $\boldsymbol{\hat
  \chi_1}\cdot \mathbf{\hat x}=-\boldsymbol{\hat \chi_2}\cdot
\mathbf{\hat x}=1$) specified at various reference times
$t_{\rm ref}/M \in [-250, -100]$.
The final kick velocity $v_k$ shows a clear sinusoidal dependence,
 as already found in, e.g., Refs.~\cite{2008PhRvD..77l4047B,
    2012PhRvD..85h4015L, Zlochower:2015wga}. The
peaks (e.g.~at $t\sim-125 M$) correspond to configurations for which
the center-of-mass velocity happens to be
at its maximum when the last oscillation is interrupted.
The orbital phase at merger can also be controlled by an overall
rotation of both spins about the orbital angular momentum. The right
panel of  Fig.~\ref{alphaseries} shows $v_k$ for binaries with spins
$\boldsymbol{\hat \chi_1} =-\boldsymbol{\hat \chi_2}=
[\cos\alpha,\sin\alpha,0]$ specified at $t_{\rm ref}=-100M$ (a similar
series of NR simulations was reported
in Ref.~\cite{2008PhRvD..77l4047B}). The right-left (left-right)
configuration
corresponds to $\alpha=0$ ($\pi$). The two curves in
Fig.~\ref{alphaseries} span the very same range, showing that the
angle $\alpha$ and the reference time $t_{\rm ref}$ are indeed
degenerate.
In practice, this means that only binaries with a specific orbital
phase at merger are subject to superkicks, thus making their
occurrence very rare. Figure~\ref{alphaprof} shows the velocity
accumulation profile for the same series of binaries with different
values of $\alpha$: the BH merger
abruptly stops the center-of-mass oscillation at different phases,
thus setting the final kick velocities.

As first noted in Refs.~\cite{2011PhRvL.107w1102L, 2013PhRvD..87h4027L},
binaries with partially aligned spins give rise to BH kicks even
larger than those imparted to binaries in the superkick
configuration.
Equal-mass, maximally spinning BH binaries are predicted to reach
$v_k\sim 5000$ km/s for spins misaligned by angles
$\theta_i=\cos^{-1}( \boldsymbol{\hat \chi_i}\cdot \mathbf{L})\sim
50^\circ$.
These recoils were dubbed \emph{hang-up kicks}, and are due to a
combination of the BH frame-dragging addition (responsible for
superkicks) and the orbital hang-up effect (which enhances the energy
radiated in GWs for aligned spins). To check that our model reproduces
these hang-up kicks, we generate $10^5$ binaries with $q=1$,
$\chi_1=\chi_2=0.8$, and isotropic spin orientations.
The largest kick detected is $v_k\sim 3300$ km/s, and is obtained for
 $\theta_1\sim \theta_2 \sim 57^\circ$.
For the same values of $q$, $\chi_1$ and $\chi_2$, the hang-up kick
fitting formula of Refs.~\cite{2011PhRvL.107w1102L,2013PhRvD..87h4027L}
returns a largest kick of $\sim 3500$ km/s  (a more careful comparison
is postponed to Sec.~\ref{statistics}).
The spacetime trajectory corresponding to one of these cases is shown
in the right panel of Fig.~\ref{centerofmass}, confirming our earlier
claims that large kicks present rather simple accumulation profiles.

\begin{figure}
\includegraphics[width=0.95\columnwidth]{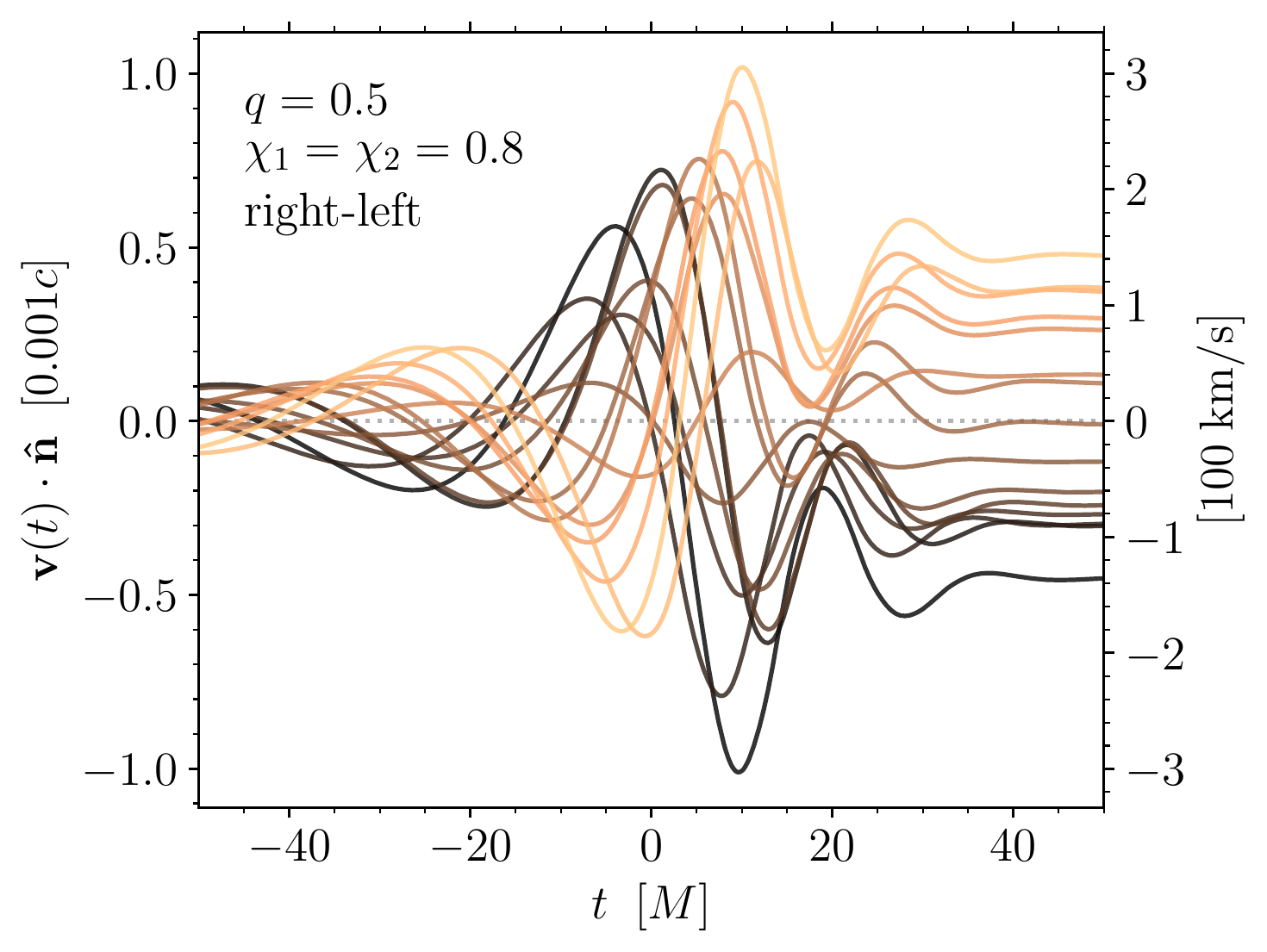}
\caption{Kick profiles for a right-left binary with $q=0.5$  and
  $\chi_1=\chi_2=0.8$ projected along various random directions
  $\mathbf{\hat n}$. Curves are colored from black to orange according
  to the final projected kick $\lim_{t\to \infty}
  \mathbf{v}(t)\cdot\mathbf{\hat n}$.}
\label{lineofsight}
\end{figure}

Finally, Fig.~\ref{lineofsight} explores projection effects of the
kick accumulation profile. For a single system with $q=0.5$ and
$\chi_1=\chi_2=0.8$ in the right-left configuration, we show the
projection of $\mathbf{v}(t)$ along various randomly chosen directions
$\mathbf{\hat n}$. Although some features are solid, the kick profile
appears rather different if viewed from different orientations. This
behavior is important to model BHs recoiling into astrophysical
environments with well-defined geometries, such as accretion
disks~\cite{2010MNRAS.401.2021R, 2010MNRAS.404..947C}, and to
implement the effect of the BH kick in waveform models through the
induced Doppler shift~\cite{2016PhRvL.117a1101G}.

\subsection{Statistical exploration and comparison with fitting formulas}
\label{statistics}

\begin{figure*}[t]
\includegraphics[width=0.847\textwidth]{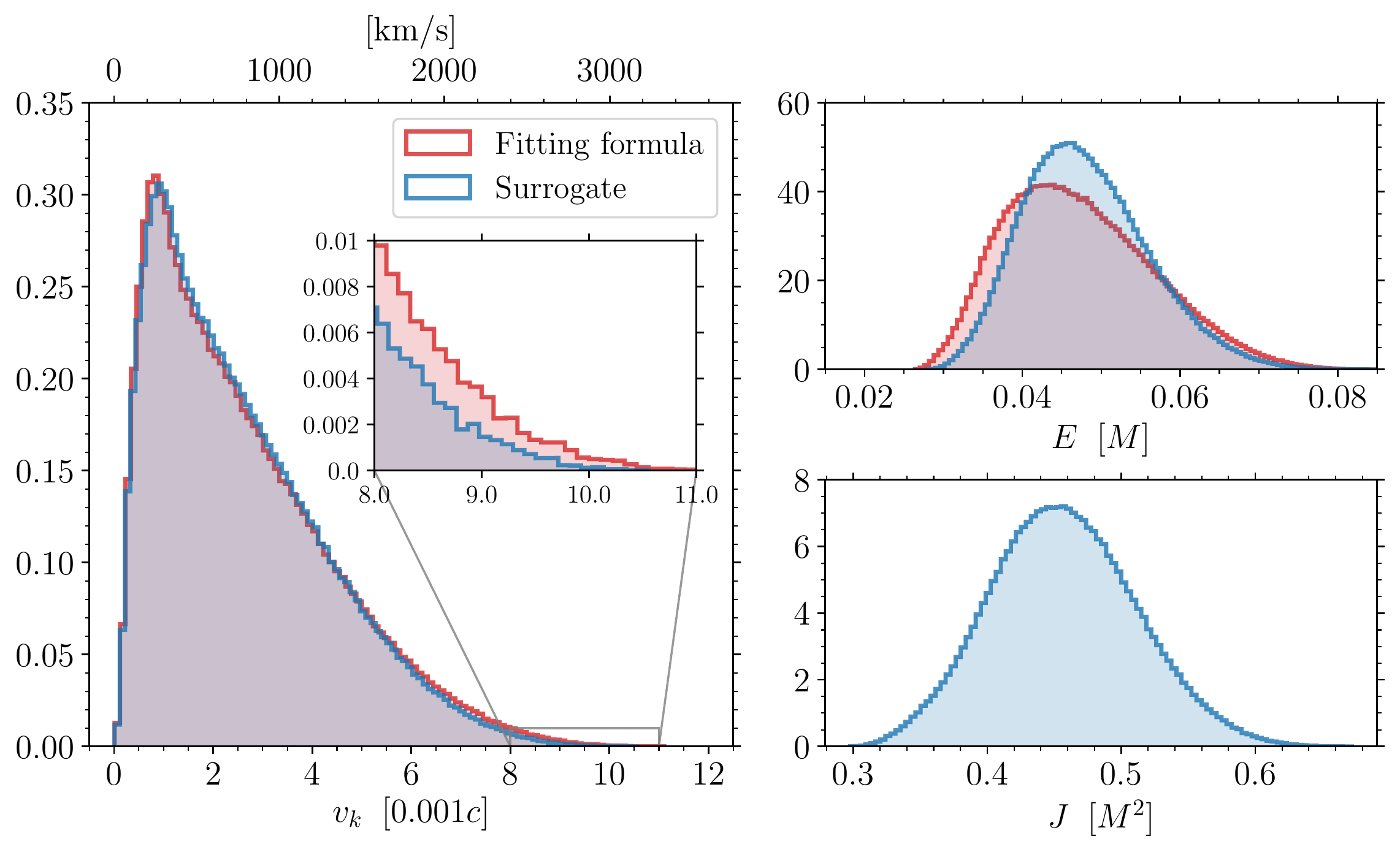}
\caption{Distribution of radiated linear momentum $v_k$ (left panel),
  energy $E$ (top right panel) and angular momentum $J$ (bottom right
  panel) for a distribution of binaries with mass ratio uniformly
  distributed in $[0.5,1]$ and spin of magnitude $\chi_i<0.8$
  uniformly distributed in volume. Our results (``Surrogate'') are
  compared to the model summarized inRef.~\cite{2016PhRvD..93l4066G} based
  on Refs.~\cite{2007ApJ...659L...5C, 2007PhRvL..98i1101G,
    2008PhRvD..77d4028L, 2012PhRvD..85h4015L, 2013PhRvD..87h4027L,
    2008PhRvD..77d4028L}~(``Fitting formula''): the two distributions
  largely agree, although differences are present for large values of
  $v_k$.}
\label{explore}
\end{figure*}

After exploring the main features of the kick profile in controlled
scenarios, we now turn our attention to statistical samples.
We generate a sample of $10^6$ binaries with mass ratio uniform in
$q\in[0.5,1]$ and spins uniformly distributed in volume with magnitude
$\chi_i\leq0.8$.
Figure~\ref{explore} shows the distributions of total energy, linear
momentum, and angular momentum radiated in GWs by this BH binary
population.
The energy and angular momentum distributions are roughly symmetric,
with peaks at $E\sim0.045 M$ and  $J\sim0.45 M^2$, respectively. The
recoil distribution peaks at $v_k\sim0.001 c$, with a long tail
extending up to $v_k\sim 0.01 c\sim 3000$ km/s. Figure~\ref{explore}
also shows predictions for $v_k$ obtained with fitting formulas
currently available in the literature. In particular, we use
the expressions summarized in Ref.~\cite{2016PhRvD..93l4066G}, which are
calibrated on various numerical simulations from Refs.~\cite{2007ApJ...659L...5C, 2007PhRvL..98i1101G,
  2008PhRvD..77d4028L, 2012PhRvD..85h4015L, 2013PhRvD..87h4027L,
  2008PhRvD..77d4028L}.  Although
kick predictions for individual binaries might differ significantly,
the two methods largely agree on the overall distribution. We note,
however, that the fitting formula tends to overestimate the number of
binaries receiving large recoils. In particular, the fractions of
binaries with $v_k>2000$ km/s are $\sim 2.4$\% and $\sim 3.2$\% for the surrogate extraction and fitting formula, respectively. The largest
kicks found in these distributions are $v_k\sim 3160$ km/s (surrogate)
and $v_k\sim 3330$ km/s (fit). We speculate that this disagreement might be
due to the calibration of the hang-up kick terms in the fitting
formula, which was only performed with $q=1$ simulations
(cf. Ref.~\cite{2015ASSP...40..185S} for a critical discussion on this
point).  Although some runs for unequal-mass binaries with largely
misaligned spins have been presented~\cite{2007ApJ...659L...5C,
  2008ApJ...682L..29B, 2009PhRvD..79f4018L, Zlochower:2015wga}, the
effect of the mass ratio on the largest kick might not be fully
captured by the expressions currently available.  Figure.~\ref{explore}
also compares the total radiated energy extracted from the surrogate
model against the final-mass fitting formula
of~\cite{2012ApJ...758...63B}, corrected according to
Eq.~(\ref{masslimits}). Agreement is found at the $\sim 2\%$ level:
the median for the surrogate (fit) estimate of $E/M$ is $\sim 0.047$
($\sim 0.046$) with standard deviations of $\sim 0.008$
($\sim 0.009$). The authors of Ref.~\cite{2017PhRvD..95f4024J} presented a careful
analysis comparing different estimates of the energy radiated
following BH mergers and reported similar, if not higher, differences
between various
approaches. %

\begin{figure}[t!]
\includegraphics[width=\columnwidth]{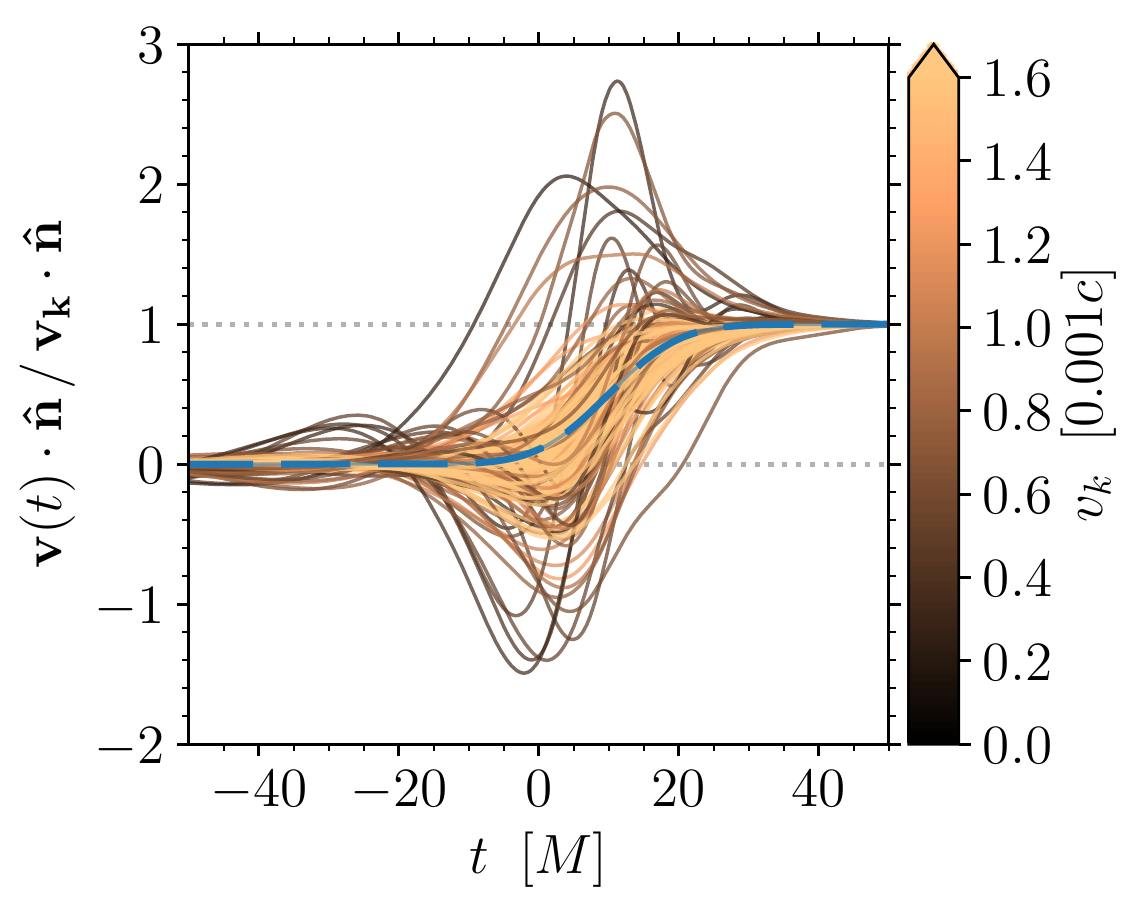}
\caption{Kick profiles $\mathbf{v}({t})$ for a sample of BH binaries
  with uniform mass ratio and isotropic spin directions projected
  along random directions $\mathbf{\hat n}$. Curves are normalized
  according to the final projected kick $\mathbf{v_k}\cdot
  \mathbf{\hat n}$ and are colored according to the total kick
  magnitude $v_k$. The dashed blue line corresponds to a Gaussian
  acceleration profile of width $\sigma=8M$ centered at $t=10M$, which
  well approximates the largest kick in our sample. Smaller kicks
  require more complicated profiles to be modeled
  carefully.
  \vspace{0.3cm}
  }
\label{normprofiles}
\end{figure}

In order to highlight the ``shape'' of the kick,
Fig.~\ref{normprofiles} shows 200 velocity accumulation profiles
$\mathbf{v}(t)$ from the same binary distribution projected along
random directions $\mathbf{\hat n}$ and normalized to the value of the
final kick $\mathbf{v_k}\cdot\mathbf{\hat n}$.
Despite the remarkable complexity explored above, the kick
accumulation profiles present very robust features. In particular,
profiles are simpler for binaries receiving large recoils, for which
the acceleration $d\mathbf{v}/dt\cdot \mathbf{\hat n}$ is well
approximated by a single Gaussian with mean $t=10M$ and width
$\sigma=8M$. Smaller kicks, on the other hand, present more
complicated profiles which typically include an
antikick~\cite{2010PhRvL.104v1101R}. These findings corroborate the
approach of Ref.~\cite{2016PhRvL.117a1101G}, where
$\mathbf{v}(t)\cdot\mathbf{\hat{n}}$ was modeled with a basis of damped
oscillatory functions.

We stress that the population explored here is far from being
astrophysically relevant.
Astrophysical processes (such as the Bardeen-Petterson effect in the case
of disk accretion~\cite{1975ApJ...195L..65B} and tidal interactions
for stellar-mass BH progenitors~\cite{1981A&A....99..126H}) deeply
modify the BH spin orientations, thus affecting the expected kick
distribution~\cite{2013MNRAS.429L..30L, 2013ApJ...774...43M,
  2015MNRAS.451.3941G}.  Moreover, PN effects in the long inspiral
before merger have been shown to preferentially suppress or enhance
recoils in specific regions of the parameter
space~\cite{2010ApJ...715.1006K, 2015MNRAS.451.3941G}.

\section{Accuracy}
\label{accuracy}

\subsection{Exploiting symmetries}
\label{exploiting}

Before presenting a detailed comparison with NR simulations, we first
perform internal tests of our kick extraction procedure by leveraging
the symmetries of the problem. For instance, equal-mass nonspinning
systems are not expected to recoil ($v_k=0$). Our extraction procedure
returns $v_k\sim 10^{-5}$, which has to be considered a numerical
error.  Following Refs.~\cite{2008PhRvL.100o1101B,2008PhRvD..78b4017B}, we
further exploit this argument using other symmetries of the system.
In particular:
\begin{enumerate}
\item[(i)] $q=1$ and $\boldsymbol{\chi_1}=\boldsymbol{\chi_2}$ imply $v_k=0$.
\item[(ii)] Aligned spins ($\boldsymbol{\chi_1}\parallel \mathbf{\hat
    L}$ and $\boldsymbol{\chi_2}\parallel \mathbf{\hat L}$) force the
  recoil to be confined to the orbital plane ($\mathbf{v_k}\cdot
  \mathbf{\hat L}=0$); this property is independent of $q$.
\item[(iii)] For $q=1$ and spins with opposite orbital-plane
  components ($\boldsymbol{\chi_1}\cdot \mathbf{\hat
    L}=\boldsymbol{\chi_2}\cdot \mathbf{\hat L}$ and
  $\boldsymbol{\chi_1}\times \mathbf{\hat L} =
  -\boldsymbol{\chi_2}\times \mathbf{\hat L}$) the kick is restricted
  to be orthogonal to the orbital plane ($\mathbf{v_k}\parallel
  \mathbf{\hat L}$).
\end{enumerate}
Some of the special cases encountered in  Sec.~\ref{anatomy} belong to
these classes. For instance, equal-mass nonspinning systems are a
trivial example of all categories. The $q=1$ up-up, down-down,
right-right and left-left cases shown in Figs.~\ref{spinaligned} and
\ref{leftright} are an instance of (i) and are therefore expected to
have $v_k=0$. All up-up, down-down, up-down and down-up configurations
are an instance of (ii), while right-left and left-right binaries with
$q=1$ are an instance of (iii).

\begin{figure}
\includegraphics[width=0.49\textwidth]{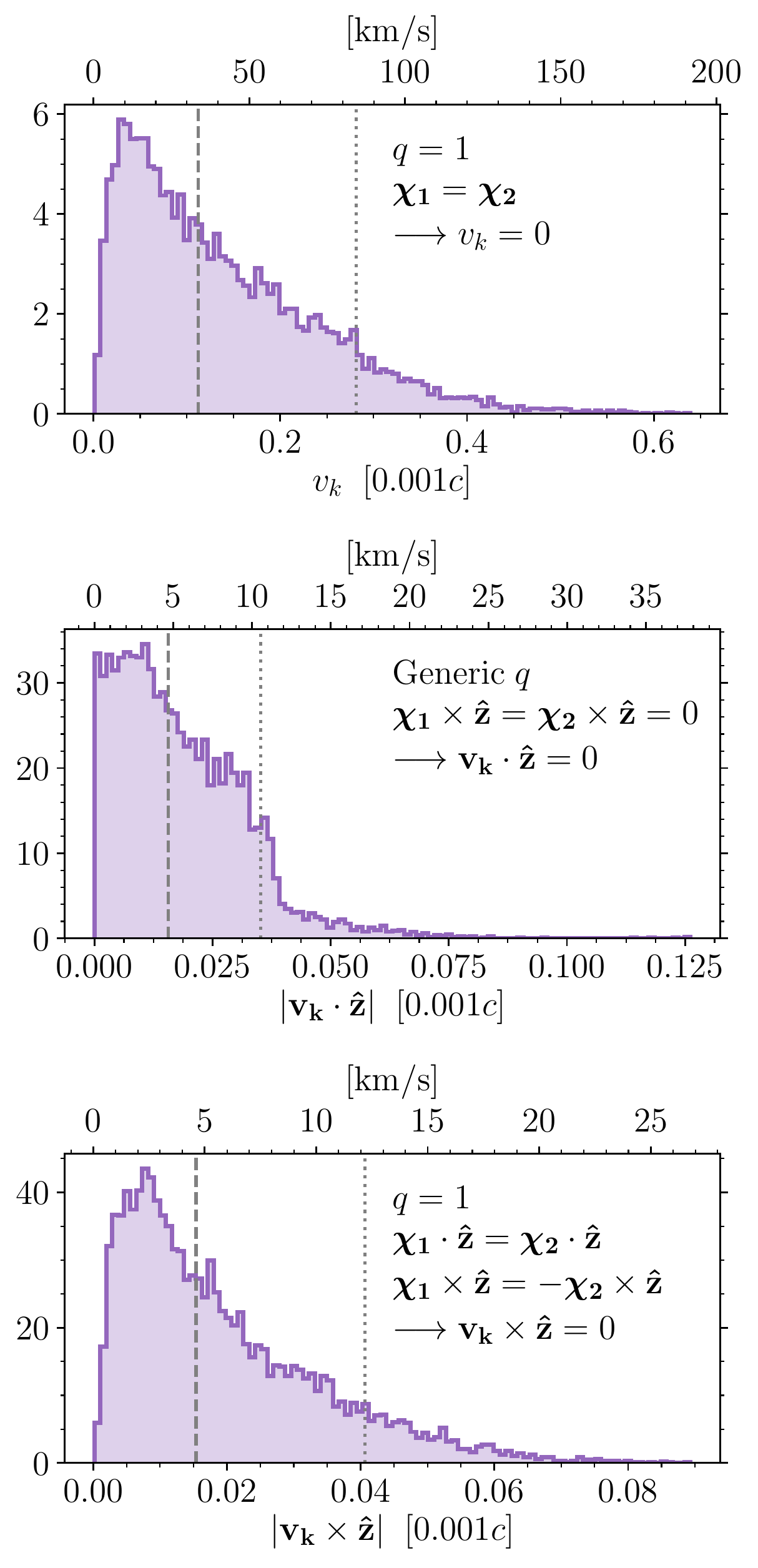}
\caption{Test of the kick numerical extraction by exploiting some of
  the symmetries of the system. All quantities shown in these plots
  are expected to be zero; deviations are interpreted as numerical
  inaccuracies of our extraction procedure. Top panel, symmetry (i):
  equal mass binaries with the same spin vectors are expected to have
  zero kicks. Middle panel, symmetry (ii): binaries with generic mass
  ratio and aligned spins are expected to have kicks in the orbital
  plane. Bottom panel, symmetry (iii): equal-mass binaries with
  opposite orbital-plane spin components and same aligned components
  are expected to have kicks directed along the binary's orbital
  angular momentum. Each panel contains a sample of $10^4$ binaries
  generated as described in the text. Dashed (dotted) lines show
  medians (90th percentiles) of the distributions.}
\label{symmetry}
\end{figure}

These symmetries are investigated in the three panels of
Fig.~\ref{symmetry}, respectively. For the top panel, we generate
binaries with $q=1$ and random spins
$\boldsymbol{\chi_1}=\boldsymbol{\chi_2}$ uniform in volume with
magnitude $<0.8$. For the middle panel, we take $q$ to be uniformly
distributed in $[0.5,1]$, generate
$\boldsymbol{\chi_i}\cdot\mathbf{\hat z}$ uniformly in $[-0.8,0.8]$,
and set all of the x and y components of the spins to zero. For the
bottom panel, we fix $q=1$, generate $\boldsymbol{\chi_1}$ uniform in
volume with magnitude $<0.8$, and set $[\chi_{2x},\chi_{2y},\chi_{2z}]
= [-\chi_{1x},-\chi_{1y},\chi_{1z}]$. The values of $v_k$,
$|\mathbf{v_k}\cdot \mathbf{\hat z}|$ and  $|\mathbf{v_k}\times
\mathbf{\hat z}|$ shown in Fig.~\ref{symmetry} are expected to be zero
under symmetries (i), (ii) and (iii), respectively.
We see that symmetry (i) exhibits the largest violations. The absolute
largest deviations are $\sim 6 \times10^{-4} c\sim 180 $ km/s, which is therefore a generous  upper limit of our numerical errors. The median of the errors is as small as  $\sim 1.1  \times10^{-4} c$, while the 90th percentile is $\sim 2.8  \times10^{-4} c$. Symmetries (ii) and (iii) are better preserved, with a precision which is roughly an order of magnitude higher. The error medians
for both are $\sim 1.5 \times10^{-5} c$.

It is worth noting that the errors reported here are rather
conservative, as they take into account inaccuracies accumulated
throughout the entire extraction pipeline---from the NR simulations
that were used to calibrate \mbox{NRSur7dq2}, to the surrogate
waveform interpolations, and finally the numerical operations described
in this paper.

\subsection{Comparison with numerical relativity simulations: SpEC}
\label{NRcomparison}

\begin{figure*}[t!]
\includegraphics[width=0.84\textwidth]{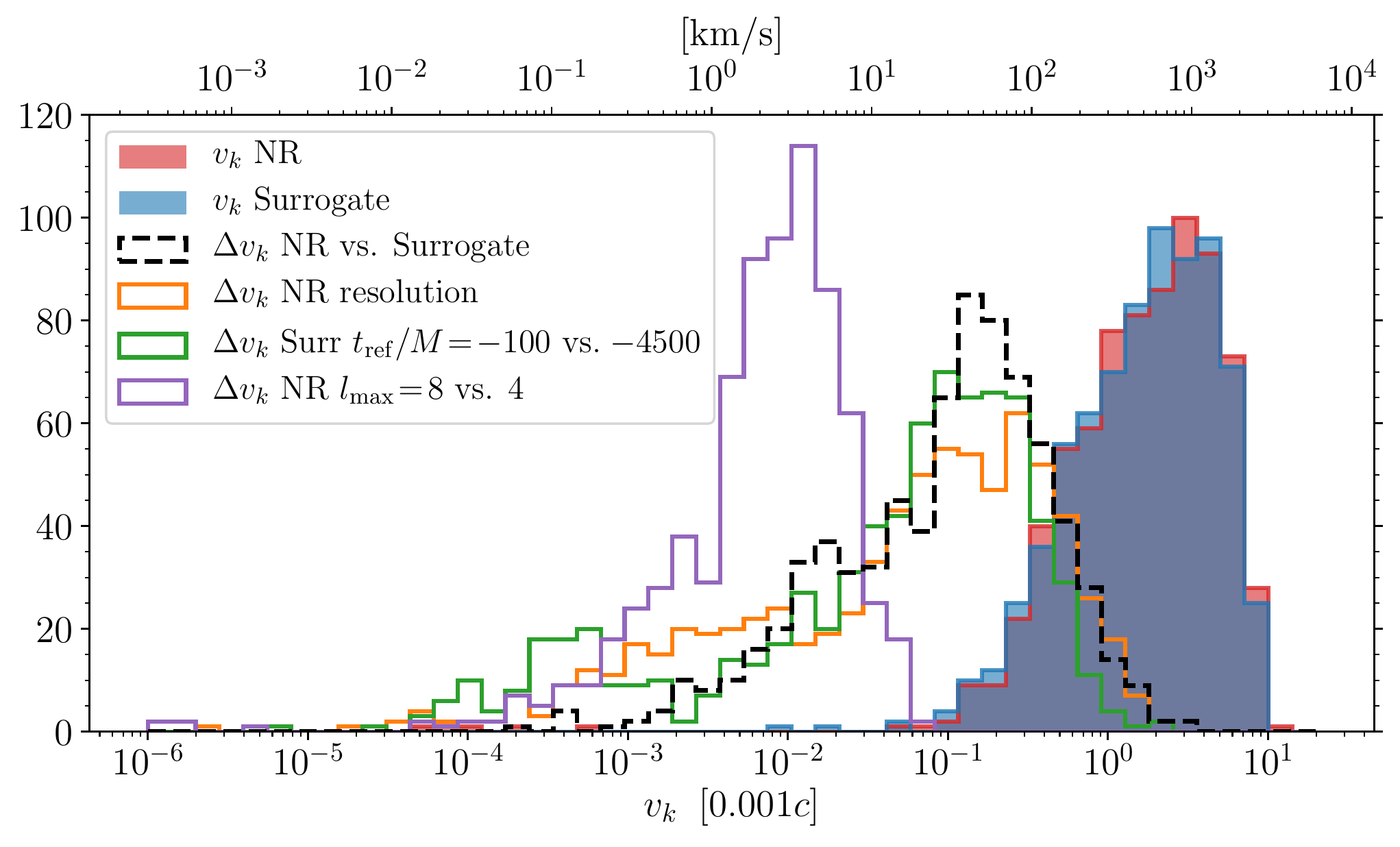}
\caption{Accuracy of the surrogate extraction of the kick velocity
  $v_k$ compared to NR simulations from SpEC. Filled histograms show
  distributions of $v_k$ extracted by both approaches, while the black
  dashed line shows residuals between the two methods. Solid thin
  lines explore some of the possible causes of the observed
  differences: the orange line shows a lower limit on the NR
  extraction accuracy, computed using the two highest resolutions
  available; the purple line shows residuals between NR kicks
  extracted with $l_{\rm max}=8$ (default) and $l_{\rm max}=4$
  (corresponding to the highest modes available in NRSur7dq2); the green
  line shows residuals in the surrogate extraction when the same NR
  runs are reproduced setting either $t_{\rm ref}=-100M$ or $t_{\rm
    ref}=-4500M$.}
\label{nrcomparison_hist}
\end{figure*}

\begin{figure}[t!]
\includegraphics[width=0.99\columnwidth]{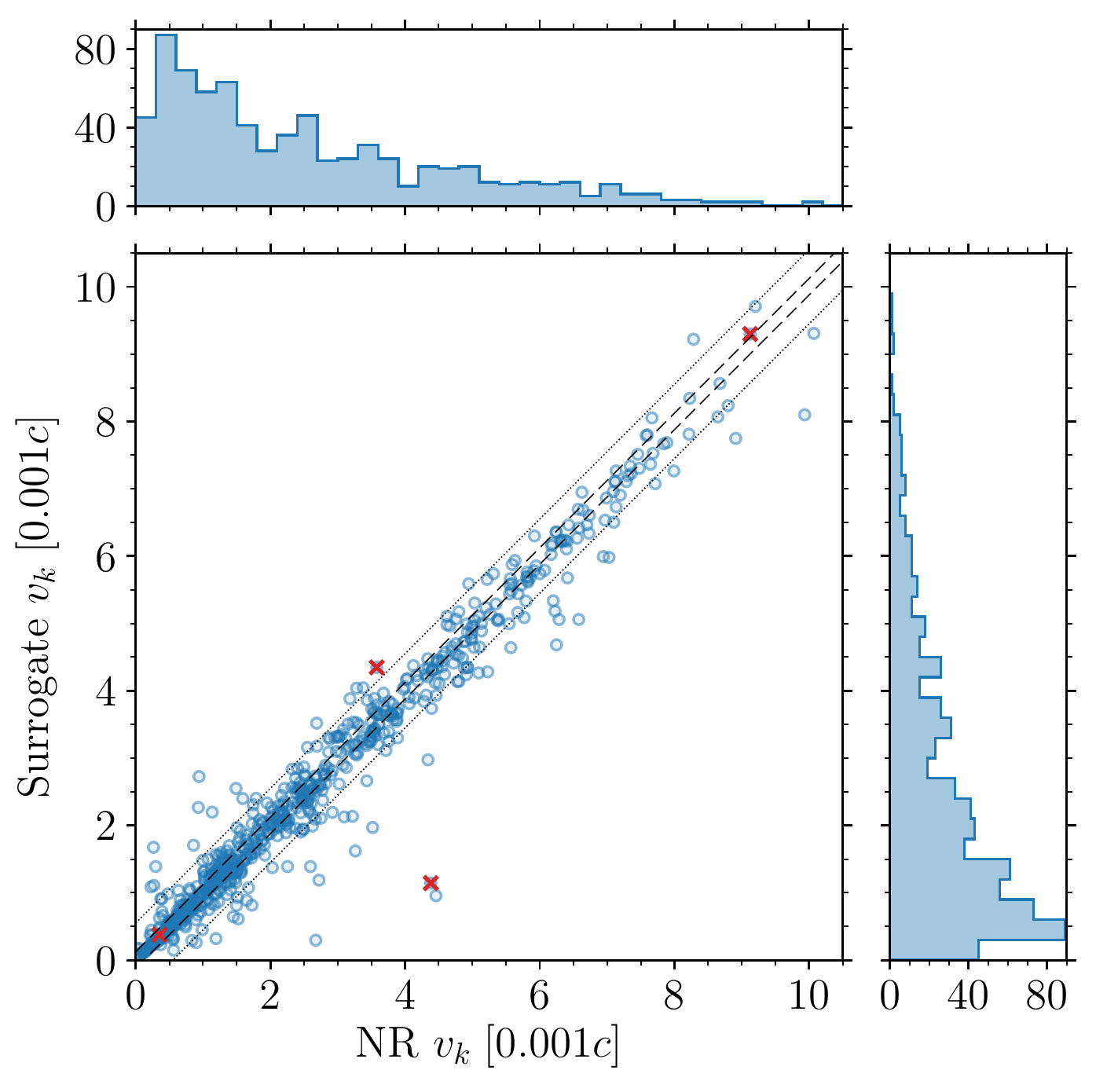}
\caption{Comparison between BH kicks extracted from NR SpEC
  simulations (horizontal) and the surrogate model \mbox{NRSur7dq2} (vertical).  The NR
  runs used here are the same that entered the surrogate model
  calibration, which was not designed to model large kicks
  specifically. 50th and 90th percentiles are shown
  with dashed and dotted lines, respectively. Red crosses mark the
  four cases explored in Fig.~\ref{nrcomparison_profiles}.}
\label{nrcomparison_scatter}
\end{figure}

\begin{figure*}[t!]
\centering
\includegraphics[width=0.85\textwidth]{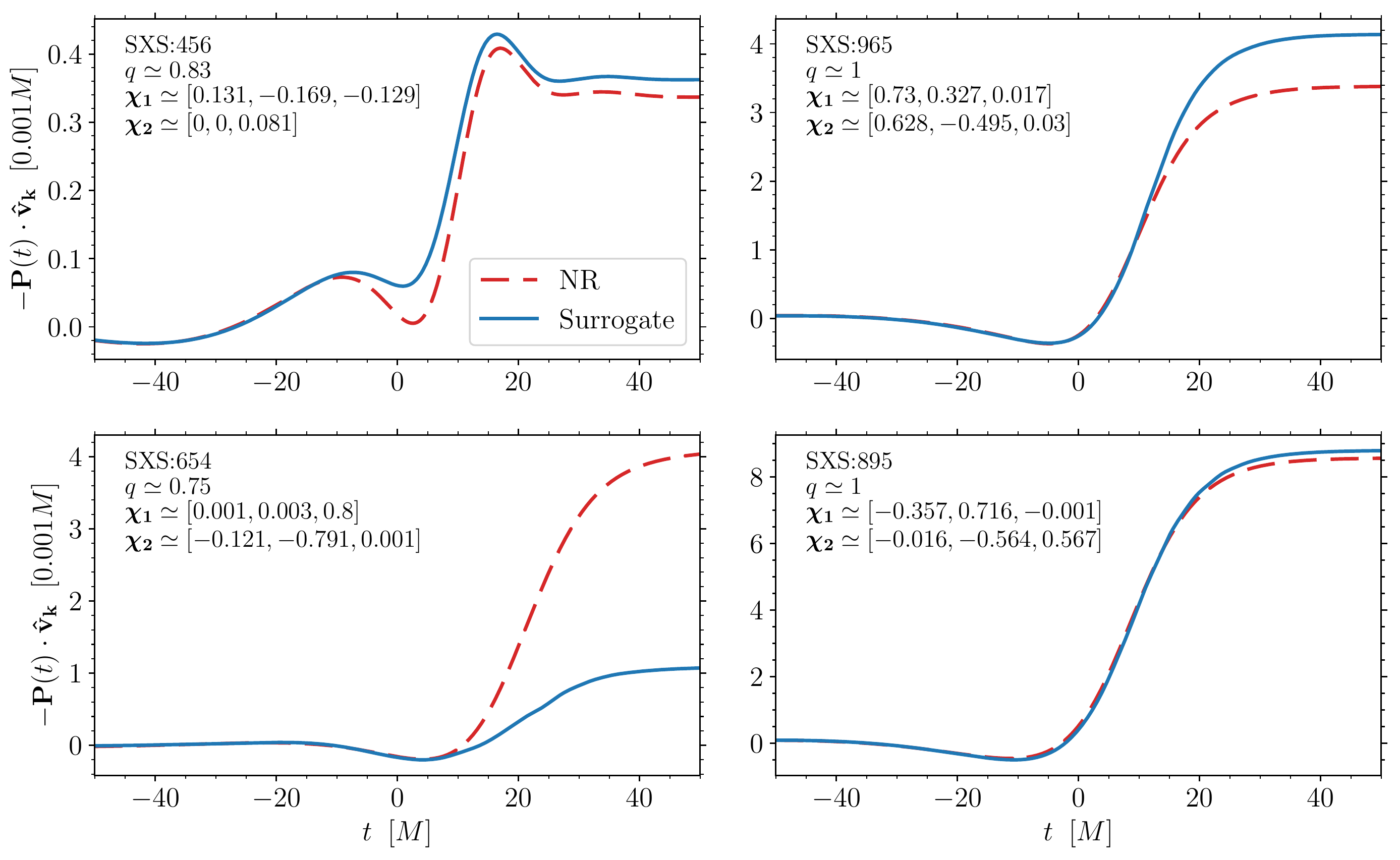}
\caption{Linear momentum profiles $\mathbf{P(t)}$ projected along the
  direction of the final kick $\mathbf{\hat v_k}$ for four selected NR
  simulations from SpEC compared to predictions obtained with the
  surrogate model. These same four cases are marked with crosses in
  Fig.~\ref{nrcomparison_scatter}. While the vast majority of the kick
  morphologies are faithfully represented, some outliers are
  present. An example is provided in the bottom right panel, where
  profiles are in good agreement before merger but then diverge at
  $t\sim10M$.}
\label{nrcomparison_profiles}
\end{figure*}

We now estimate the accuracy of our extraction procedure by directly
comparing our results to numerical relativity simulations from the
SpEC code~\cite{2000PhRvD..62h4032K}.
In particular, we compare against the 744
simulations\footnote{%
  NRSur7dq2 is trained on 886 waveforms obtained from 744 simulations
  --- 142 simulations have $q=1$ and
  $\boldsymbol{\chi_1} \ne \boldsymbol{\chi_2}$, so that a rotation
  enables one simulation to represent two sets of binary parameters
  and therefore two input waveforms~\cite{2017PhRvD..96b4058B}.}
used to construct NRSur7dq2~\cite{2017PhRvD..96b4058B}. These
simulations constitute the majority of the waveforms available in the
SpEC catalog~\cite{2013PhRvL.111x1104M}
in the relevant parameter range, and especially so for generic spin
orientations.
This is not the most ideal comparison: each of these numerical
simulations occupies a special point in the binary parameter space of
the surrogate model.
However, it is worth noting that (i)~the surrogate waveforms do not
reproduce the NR waveforms exactly, even at the parameter-space
location of the simulations that entered the training process; and
(ii)~NRSur7dq2 was designed to maximize the overlap between the
interpolated and the NR strain $h$, not to accurately model BH kicks.
The comparison to NR simulations will therefore be sensitive to errors
from the surrogate's reproduction of the training set of gravitational
waveforms, but insensitive to errors from the surrogate's
interpolation between these waveforms.

Recoils are extracted from SpEC waveforms using the expressions
reported in Sec.~\ref{radiatedexpressions},
and normalized by the remnant mass computed from the BH horizon at
the end of the SpEC simulation.
We include modes up to $l_{\rm max}=8$ from the highest-resolution data.
To compare with the surrogate kick, we must determine the correct
binary parameters by first time shifting and rotating the NR waveforms
consistently with NRSur7dq2 (per criteria given in
Sec.~\ref{surrogatemodels}) and then measuring the BH spins at $t_{\rm
  ref}=-4500M$ as in~\cite{2017PhRvD..96b4058B}. Consequently the
surrogate is evaluated with $t_{\rm ref} = -4500 M$.
Filled histograms in Fig.~\ref{nrcomparison_hist} show the
distributions of $v_k$ obtained for both the NR and surrogate
extractions. Differences $\Delta v_k$ between the two (thick dashed
line) are typically $\sim 10^{-4} c$; 90\% of the simulations are
reproduced within $\Delta v_k = 5.5 \cdot 10^{-4} c$.
In this histogram we also plot several sources of error to evaluate
their importance.
One of these is the difference between NR kicks extracted from
different resolutions of each SpEC simulation ---a solid upper limit
on the accuracy of the NR kick extraction.
This also presents a tail up to $\sim 2\cdot 10^{-3} c$, similar to
that of $\Delta v_k$.
The selection of the reference time $t_{\rm ref}$ in the surrogate
extraction is a marginally smaller effect, with tail up to $\sim
10^{-3} c$.
The contribution of higher-order modes $l>4$ to the NR kick is a
subdominant effect and contributes only on the scale of $\sim10^{-5}
c$.
Finally, the error from evaluating the kick at a finite time $t=100M$,
instead of taking the kick's $t\to\infty$ limit, is negligible: the
NR kicks extracted at $t=100M$ and $135M$ (each simulation has a
different final time in $[139M,165M]$) differ by $\sim 10^{-8}c$
only.
The surrogate-to-NR comparison is also presented as a scatter plot in
Fig.~\ref{nrcomparison_scatter}, which shows how the surrogate kick
extraction faithfully reproduces the vast majority of the
simulations. A few outliers with $\Delta v_k\sim 2\cdot 10^{-3} c$ are
present in the bottom-center panel of the figure (also in
Fig.~\ref{nrcomparison_hist} as the tail of the $\Delta v_k$
distribution), for which our surrogate extraction
underestimates the value of $v_k$.
These are cases where the surrogate model fails to correctly reproduce
some cycles in the waveform's higher harmonics around the time of
merger, when the majority of the kick is being accumulated.
We note that cases with large $\Delta v_k$ are preferentially located
at the high-spin edge of the NRSur7dq2 parameter space:
the three outliers mentioned above, and $\sim 2/3$ among the 5\% of
cases with the largest $\Delta v_k$,
have $\chi_1 = \chi_2 = 0.8$.
This occurs because the error
of the SpEC simulations, and consequently the surrogate model waveforms,
increases towards this maximum-spin boundary.
Restricting to the 464 NR simulations (or $\sim 2/3$ of the sample)
with zero or one spin of magnitude $\chi = 0.8$, we find that the surrogate
reproduces 90\% of the kicks within $\Delta v_k$ of
$3.8\cdot10^{-4}c \sim 113$ km/s.
The error is about twice as large for the 280 simulations (or $\sim
1/3$ of the sample) with $\chi_1 = \chi_2 = 0.8$, with 90\% of the
kicks being within
$7.7\cdot10^{-4}c \sim 232$ km/s.

Finally, Fig.~\ref{nrcomparison_profiles} shows comparisons for the
kick accumulation profiles
$\mathbf{P}(t)\cdot \mathbf{\hat  v_k}$
in four selected cases. We find the the surrogate model reproduces not
only the kick magnitude $v_k$, but also the morphology of the time
accumulation profile for the vast majority of the NR simulations. The
lower left
panel of Fig.~\ref{nrcomparison_profiles} shows one of the few
outliers, which has $\Delta v_k\sim 3 \cdot 10^{-3} c$. The NR
and surrogate profiles diverge around $t \sim 10M$, when the surrogate
fails to capture the merger waveform.
These two curves appear similar to the kick profiles of
Fig.~\ref{alphaprof}, suggesting that the surrogate model fails to
reconstruct the orbital phase at merger. Even if \mbox{NRSur7dq2} well
reproduces the strain $h$, its small errors might propagate to the
phase of center-of mass-oscillation causing a relatively large error
on the final kick velocity.

\subsection{Comparison with numerical relativity simulations: LazEv}

\begin{figure}
\includegraphics[width=\columnwidth]{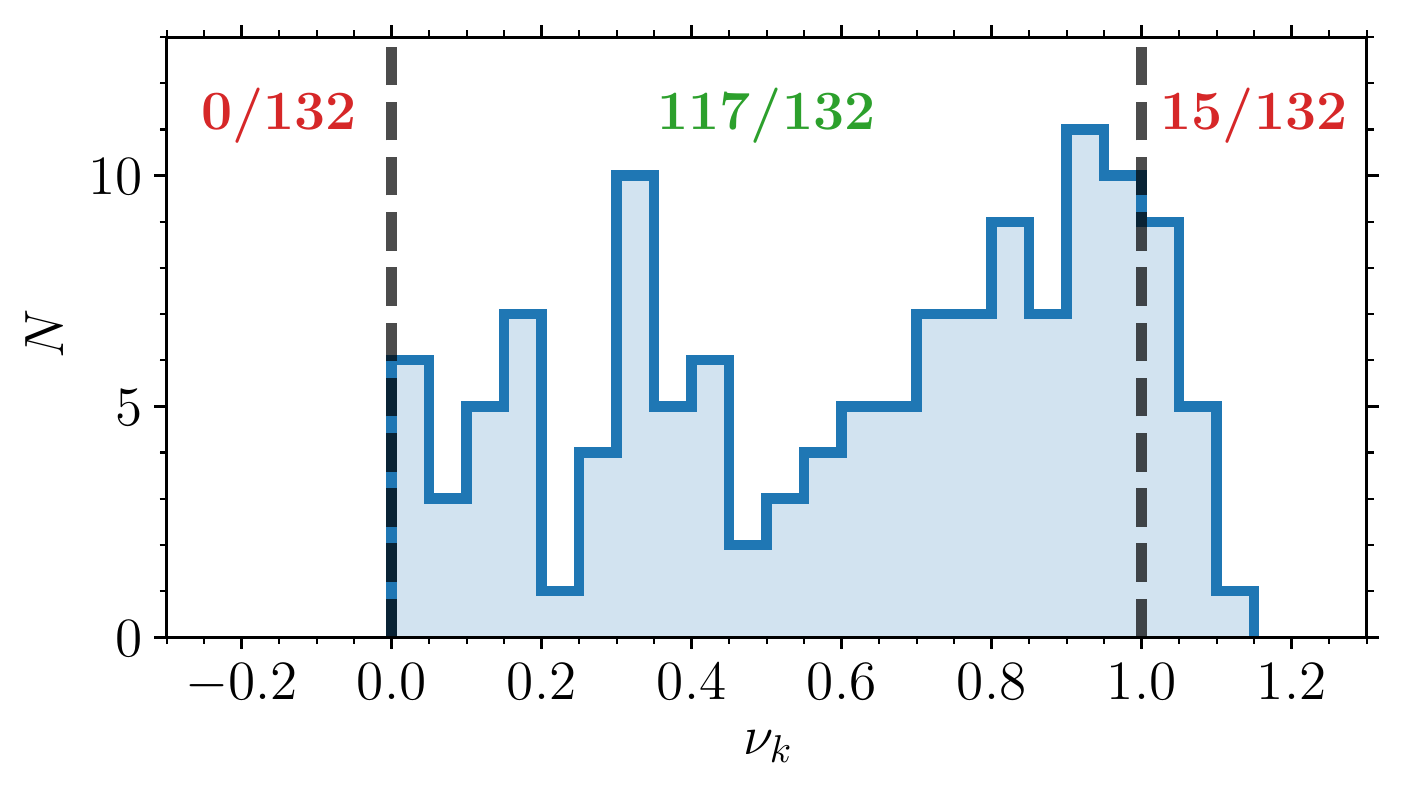}
\caption{Distribution of BH kicks extracted from 132 NR LazEV simulations \cite{2013PhRvD..87h4027L,Zlochower:2015wga}, rescaled between the minimum and maximum kicks obtained from \mbox{NRSur7dq2} [cf. Eq.~(\ref{nuk})]. If $0\leq \nu_k \leq 1$, there exists a suitable choice of $t_{\rm ref}$ for which the surrogate model reproduces the NR value of the kick. On the other hand, the NR data cannot be reproduced if $\nu_k<0$ or $\nu_k>1$.}
\label{RIT_check}
\end{figure}

Finally, we compare our results against NR simulations performed by the RIT group with the LazEv code \cite{2005PhRvD..72b4021Z}. This additional comparison is noteworthy because not only were these simulations not used in the surrogate calibration, but they were performed with a completely different numerical scheme (for a detailed comparison between SpEC and LazEv see Ref.~\cite{2016CQGra..33x4002L}).

We compare against several series of simulations performed by Lousto and Zlochower that vary over the relative azimuthal projection of the spin (i.e.~the angle $\alpha$ defined in Sec.~\ref{anatomy}) \cite{2013PhRvD..87h4027L,Zlochower:2015wga}.
Of the 223 NR simulations described in these references, 132 of them lie within the parameter range covered by NRSur7dq2.\footnote{Some of the simulations have parameters which exceed the range of validity of the surrogate model only very marginally ($q\simeq 0.498$ and/or $\chi_i\simeq 0.802$). We do not filter those runs out, but rather use NRSur7dq2 in extrapolation mode.} We extract horizon masses, spins, and final kicks from the relevant tables in Refs.~\cite{2013PhRvD..87h4027L,Zlochower:2015wga}; then, we use the mass ratios and spins as inputs to NRSur7dq2.
Case-by-case comparisons between the RIT simulations and the surrogate model are not possible because differences in gauges preclude us from converting their initial separations to our $t_{\mathrm{ref}}$'s.
We can, however, check for each case whether there exists a choice of $t_{\mathrm{ref}}$ for which the surrogate reproduces the reported value of the kick.

To this end, we rescale each of the RIT kick values $v_k^{(\rm NR)}$ with an affine transformation determined by the minimum and maximum surrogate kicks $v_k^{(\rm surr)}$ as $t_{\mathrm{ref}}$ is varied over the range $t_{\rm ref}/M\in [-4500,-100]$, while holding all other parameters fixed: %
\begin{equation}
\nu_k = \frac{ v_k^{(\rm NR)} - \min_{t_{\rm ref}}\, v_k^{(\rm surr)}}{ \max_{t_{\rm ref}}\, v_k^{(\rm surr)} - \min_{t_{\rm ref}}\,v_k^{(\rm surr)}}
\,.
\label{nuk}
\end{equation}
Therefore the kicks from Refs.~\cite{2013PhRvD..87h4027L,Zlochower:2015wga} that can be reproduced lie in the range $0\le\nu_{k}\le 1$.
The resulting distribution of $\nu_k$ is shown in Fig.~\ref{RIT_check}. We find that $0\leq \nu_k\leq 1$ for $117/132\simeq 89\%$ of the simulations. The remaining simulations cannot be matched by our procedure; in particular, the surrogate underestimates the NR result in $15/132\simeq 11\%$ of the cases for which $\nu_k>1$ (no simulations are found with $\nu_{k}<0$). We stress, however, that these disagreements are very moderate, with $\nu_k<1.12$ 
over all the simulations we analyzed.

The different comparisons presented in this section show that the
surrogate kick extraction reaches precisions similar to those of the
NR simulations that entered its
calibration, well respects the symmetries of the problem, and matches kick results obtained with an independent NR code. We quote
an overall average precision of $40$ km/s on the surrogate extraction
of $v_k$.

\addtolength{\tabcolsep}{+13pt}
\begin{table*}[htp]
\begin{center}
\begin{tabular}{llll}
\hline \hline
Method & Description & Equation & Default inputs
\\
\hline \hline
\texttt{sur()} & Instance of the surrogate class from \mbox{NRSur7dq2}.
&&\\
\texttt{q} & Binary mass ratio $q\in[0.5,1]$. && $q=1$.\\
\texttt{chi1} & Spin vector $\boldsymbol{\chi_1}$ of the heavier BH at $t_{\rm ref}$. && $\boldsymbol{\chi_1}=[0,0,0]$.\\
\texttt{chi2} & Spin vector $\boldsymbol{\chi_2}$ of the lighter BH at $t_{\rm ref}$. &&$\boldsymbol{\chi_2}=[0,0,0]$.\\
\texttt{t\_ref} & Reference time $t_{\rm ref}/M\in[-4500,-100]$. && $t_{\rm ref}/M=-100$.\\
\texttt{times} & Time nodes $t_i/M\in[-4500,100]$. &&\\
\texttt{lmax} & Largest available $l$ mode ($l_{\rm max}=4$ in \mbox{NRSur7dq2}).&&\\
\texttt{h(l,m)} & Modes of the complex GW strain $h^{lm}$.&Eq.~(\ref{hmodes})& \\
\texttt{hdot(l,m)} & Modes of the time derivative $\dot h^{lm}$ && \\
\texttt{dEdt} & Energy flux $dE/dt$. & Eq.~(\ref{energyflux})& \\
\texttt{Eoft} & Radiated energy profile $E(t)$. & & \\
\texttt{Erad} & Total radiated energy $\lim_{t\to\infty}E(t)$. & & \\
\texttt{Moft} & Mass profile $M(t)$. & Eq.~(\ref{Moft})& \\
\texttt{Mrad} & Mass of the remnant BH $\lim_{t\to\infty}M(t)$. & & \\
\texttt{Mfin} & Mass of the remnant BH in units of the mass at $t=-\infty$. &Eq.~(\ref{masslimits}) & \\
\texttt{dPdt} & Linear momentum flux $d\mathbf{P}/dt$ & Eqs.~(\ref{eq:dt_px}-\ref{eq:dt_pz})& \\
\texttt{Poft} & Radiated linear momentum profile $\mathbf{P}(t)$. & & \\
\texttt{Prad} & Total radiated linear momentum $\lim_{t\to\infty}|\mathbf{P}(t)|$. & & \\
\texttt{voft} & Recoil velocity profile $\mathbf{v}(t)$. & Eq.~(\ref{voftprofile})& \\
\texttt{kickcomp} &  Kick velocity, vector $\mathbf{v_k}=\lim_{t\to\infty}\mathbf{v}(t)$. & Eq.~(\ref{vkicklimit})& \\
\texttt{kick} &  Kick velocity, magnitude $v_k$. & & \\
\texttt{kickdir} &  Kick velocity, unit vector $\mathbf{\hat v_k}= \mathbf{v_k}/v_k$. & & \\
\texttt{dJdt} & Angular momentum flux $d\mathbf{J}/dt$. & Eqs.~(\ref{eq:dt_jx}-\ref{eq:dt_jz})& \\
\texttt{Joft} & Radiated angular momentum profile $\mathbf{J}(t)$. & & \\
\texttt{Jrad} & Total radiated angular momentum $\lim_{t\to\infty}|\mathbf{J}(t)|$. & & \\
\texttt{xoft} & Center-of-mass  trajectory $\mathbf{x}(t)=\int \mathbf{v}(t) dt$. & & \\
\hline \hline
\end{tabular}
\end{center}
\caption{Main methods of the \texttt{surrkick} class. A class
  instance has to be initialized with
  e.g.~\texttt{sk=surrkick.surrkick(q=1,chi1=[0,0,0],chi2=[0,0,0],t\char`_ref=-100)}.
  Methods can then be accessed with e.g.~\texttt{sk.voft}.}
\label{codefunctions}
\end{table*}
\addtolength{\tabcolsep}{-13pt}

\section{Code distribution and usage}
\label{code}

Our numerical code, \textsc{surrkick}, is publicly available as a
module for the Python programming language. The latest stable release
is kept updated on the Python Package Index (PyPI) and can be
installed via
\begin{verbatim}
    pip install surrkick
\end{verbatim}
Python packages {numpy}~\citep{Walt}, {scipy}~\cite{Jones:2001aa},
{matplotlib}~\citep{2007CSE.....9...90H}, {h5py}~\cite{h5py},
{pathos}~\cite{mckerns-proc-scipy-2011},
tdqm~\cite{casper_da_costa_luis_2017_1012577},
{\mbox{NRSur7dq2}}~\cite{2017PhRvD..96b4058B} and
precession~\cite{2016PhRvD..93l4066G}
are specified as dependencies and will automatically be installed if
missing.
The \textsc{surrkick} module has to be imported with
\begin{verbatim}
    import surrkick
\end{verbatim}
from within a Python environment. Information on all classes, methods,
and functions of the code can be obtained from the code docstrings
using Python's \texttt{help} function. \textsc{surrkick} is hosted
under version control on GitHub at
\href{https://github.com/dgerosa/surrkick}{github.com/dgerosa/surrkick},
where development versions are available. Further information and code
outputs can be found at
\href{https://davidegerosa.com/surrkick}{davidegerosa.com/surrkick}.
\textsc{surrkick} is structured as an add-on to any waveform
approximant. In particular, it will be straightforward to update it as
new surrogate models become available.
The code is currently compatible with Python 2; porting to Python 3 is
foreseen.  Results in this paper were obtained with version 1.1 of
\textsc{surrkick}.

All of the main functionalities of the code are provided as methods of a
single class \texttt{surrkick.surrkick}. An instance of the class is
created providing mass ratio $q$, spin vectors $\boldsymbol\chi_{i}$
and  reference time $t_{\rm ref}/M$:
\begin{verbatim}
    sk=surrkick.surrkick(q=1,chi1=[0,0,0],
                 chi2=[0,0,0],t_ref=-100)
\end{verbatim}
A list of the relevant methods is provided in Table~\ref{codefunctions}.
All quantities are returned in units of the binary's total mass
(i.e.~$c=G=M=1$). Time profiles are evaluated at the time nodes
\texttt{sk.times}. For instance, the following code snippet computes
the final kick imparted to a right-left binary with $q=0.5$ and
$\chi_1=\chi_2=0.8$, and plots the velocity profile $\mathbf{v}(t)$
projected along $\mathbf{\hat x}$, $\mathbf{\hat y}$, $\mathbf{\hat
  z}$ and $\mathbf{\hat v_k}$.

\begin{samepage}
\begin{verbatim}
    import surrkick
    import matplotlib.pyplot as plt
    sk=surrkick.surrkick(q=0.5,chi1=[0.8,0,0],
        chi2=[-0.8,0,0])
    print "vk/c=", sk.kick
    plt.plot(sk.times,sk.voft[:,0],label="x")
    plt.plot(sk.times,sk.voft[:,1],label="y")
    plt.plot(sk.times,sk.voft[:,2],label="z")
    plt.plot(sk.times,surrkick.project(sk.voft,
        sk.kickdir),label="vk")
    plt.xlim(-100,100)
    plt.legend()
    plt.show()
\end{verbatim}
\end{samepage}

The class \texttt{surrkick.plots} provides tools to reproduce all
figures and results presented in this paper. The snippet above is
implemented as \texttt{surrkick.plots.minimal()}.

Performance of the code was evaluated on a single processor of an
Intel Xeon CPU E5-2660 v3 @2.60GHz averaging over $10^3$ binaries with
generic parameters.  Computation of $v_k$ takes $\sim 0.1$ s, where
$\sim50$ ms are spent evaluating $h$ from
\mbox{NRSur7dq2}~\cite{2017PhRvD..96b4058B} and $\sim 50$ ms are spent
integrating the energy and linear momentum fluxes. These low
execution times make our code ideal to be ported into large-scale
computational studies.

\section{Conclusions}
\label{conclusions}

New waveform approximants able to model precessing BH binaries with
higher harmonics have been recently developed for GW detection and
parameter estimation. Here we have shown, for the first time, how these
tools present an interesting by-product, namely the quick and reliable
estimation of energy and momenta radiated in GWs during BH inspirals
and mergers.
In particular, the dissipation of linear momentum is responsible for
powerful BH recoils, which might even eject BHs from their host
galaxies.
We exploited the recent NR surrogate model
\mbox{NRSur7dq2}~\cite{2017PhRvD..96b4058B} to explore the
phenomenology of the recoil
velocity profile $\mathbf{v}(t)$ imparted to generic binaries as
they merge. Our findings are implemented in the numerical code
\textsc{surrkick}, which is made available to the community as a
module for the Python programming language.

Our extraction procedure inherits both strengths and weaknesses of
\mbox{NRSur7dq2}. The model can reproduce the GW strain with
mismatches $\sim 10^{-3}$, orders of magnitude better than any other
model currently available. This translates into an average accuracy
$\Delta v_k/c \lesssim 10^{-4}$  on the recoil estimates. The model
has only been calibrated on BH binaries with mass ratios $q\geq0.5$
and spin magnitudes $\chi_i\leq0.8$. Both \mbox{NRSur7dq2} and
\textsc{surrkick} can in principle be used outside this range, but
those extrapolations have not been tested accurately.
\mbox{NRSur7dq2} provides evolutions over a time $\Delta t\sim 5000M$,
corresponding to $\sim 20$ orbits before merger. While this is a
severe limitation for waveform modeling (because low-mass systems
spend many more cycles in the sensitivity windows of the detectors),
it is irrelevant for kick estimation. Linear momentum emission is
concentrated in a small time window ($2 \sigma \sim 20M$) around
merger which is well covered by \mbox{NRSur7dq2}.

The tools presented here provide an alternative way to estimate BH
kicks which, contrary to fitting formulas, does not require specific
ans\"{a}tze. Moreover, they provide information on the full
$\mathbf{v}(t)$ profile, not just the final recoil velocity
$v_k$. With executions times of $\sim 0.1$ s, our approach allows for
quick and reliable implementations of BH kicks in a variety of
astrophysical studies, from galaxy evolution codes to population
synthesis studies of compact binaries.
Future developments include building new NR surrogate models
specifically designed to accurately reproduce mass, spin, and recoil of
the post-merger BH.

\bigskip
\acknowledgments
We thank Jonathan Blackman, Chad Galley, Mark Scheel, Ulrich Sperhake,
Saul Teukolsky, and Vijay Varma for fruitful discussions and technical
help.
D.G. is supported by NASA through Einstein Postdoctoral Fellowship
Grant No.~PF6-170152 awarded by the Chandra X-ray Center, which is operated by the
Smithsonian Astrophysical Observatory for NASA under Contract
NAS8--03060.
F.H. acknowledges the support of the Sherman Fairchild Foundation, and NSF grants PHY-1404569, PHY-1708212, and PHY-1708213 at Caltech.
L.C.S. acknowledges the support of NSF grant PHY-1404569 and the Brinson
Foundation.
Computations were performed on resources provided by NSF CAREER Award PHY-1151197, and on the Wheeler cluster at Caltech, which is supported by the Sherman Fairchild Foundation and by Caltech.

\bibliographystyle{apsrev4-1}
\bibliography{surrkick}

\end{document}